\documentclass[11pt, a4paper]{article}

\usepackage{setspace}
\usepackage{amssymb}
\usepackage{cite}
\usepackage{amsmath}
\usepackage{dsfont}
\usepackage{amsfonts}
\usepackage{epsfig}
\usepackage{latexsym}
\usepackage{slashed}
\usepackage{xcolor}
\usepackage{cancel}
\usepackage{verbatim}
\usepackage{nicefrac}
\usepackage{amsbsy}
\usepackage{graphicx}
\usepackage[heightrounded]{geometry}
\usepackage{comment}

\allowdisplaybreaks[1]

\def\nb{n_B}
\def\nbb{n_{\bar{B}}}
\def\ng{n_\gamma}

\newcommand{\rmeq}{{\rm eq}}
\newcommand{\barphi}{\bar{\phi}}
\newcommand{\barell}{\bar{\ell}}

\newcommand{\bh}[1]{\hat{\bf #1}}

\newcommand{\rmd}{{\rm d}}
\newcommand{\rmi}{{\rm i}}
\newcommand{\rme}{{\rm e}}
\newcommand{\tr}{{\rm tr}}

\newcommand{\CP}{{C\!P}}
\newcommand{\hc}{{\rm h.c.}}
\newcommand{\gb}{\boldsymbol{\gamma}}

\newcommand{\rmIm}{{\rm Im}}
\newcommand{\D}{\Delta}
\newcommand{\Dt}{\tilde{\Delta}}

\def\openone{\leavevmode\hbox{\small1\kern-3.8pt\normalsize1}}
\def\a{\alpha}
\def\b{\beta}
\def\c{\chi}
\def\d{\delta}
\def\e{\epsilon}
\def\f{\phi}
\def\g{\gamma}
\def\h{\eta}

\def\l{\lambda}

\def\o{\omega}
\def\p{\pi}
\def\q{\theta}

\def\D{\Delta}

\def\G{\Gamma}

\def\P{\Pi}

\def\S{\Sigma}

\def\cl{{\cal L}}

\def\bo{{\raise-.3ex\hbox{\large$\Box$}}}               % D'Alembertian
\def\face{{\raise.2ex\hbox{$\displaystyle \bigodot$}\mskip-2.2mu \llap {$\ddot
        \smile$}}}                                      % happy face
\def\leftrightarrowfill{$\mathsurround=0pt \mathord\leftarrow \mkern-6mu
        \cleaders\hbox{$\mkern-2mu \mathord- \mkern-2mu$}\hfill
        \mkern-6mu \mathord\rightarrow$}       % <--> double differential
\def\dvec#1{\vbox{\ialign{##\crcr
        \leftrightarrowfill\crcr\noalign{\kern-1pt\nointerlineskip}
        $\hfil\displaystyle{#1}\hfil$\crcr}}}           % <--> accent
\def\beq{\begin{equation}}
\def\eeq{\end{equation}}

\def\beqx{\begin{displaymath}}
\def\eeqx{\end{displaymath}}

\def\bea{\begin{eqnarray}}
\def\eea{\end{eqnarray}}

\def\bs{\begin{split}}
\def\ensp{\end{split}}

\def\ba{\begin{align}}
\def\ea{\end{align}}

\def\bef{\begin{figure}}
\def\eef{\end{figure}}

\def\bec{\begin{center}}
\def\eec{\end{center}}

\begin{document}

\begin{flushright}
MPP-2011-61\\[7mm]
\end{flushright}

\begin{center}
  {\bf \Large Hard-Thermal-Loop Corrections in Leptogenesis I:\\ $\boldsymbol{C\!P}$-Asymmetries}\\[4mm]

%\today\\[4mm]

Clemens P.~Kie\ss ig%$^\star$
\footnote{E-mail: \texttt{ckiessig@mpp.mpg.de}},
%Alois S.~Kabelschacht
%\footnote{E-mail: \texttt{kabel@mpp.mpg.de}} ,
Michael Pl\"umacher%$^\star$
\footnote{E-mail: \texttt{pluemi@mpp.mpg.de}}
%Markus H.~Thoma$^\dagger$
%\footnote{E-mail: \texttt{mthoma@mpe.mpg.de}}, 

\vspace*{0.5cm}
%$^\star$ 
\it
Max-Planck-Institut f\"{u}r Physik (Werner-Heisenberg-Institut),\\
F\"ohringer Ring 6, D-80805 M\"unchen, Germany\\[0.1cm]
%$^\dagger$ \it Max-Planck-Institut f\"ur extraterrestrische Physik,\\
%Giessenbachstra\ss e, D-85748 Garching, Germany

\vspace*{0.4cm}
\end{center}

\begin{abstract}

  We investigate hard-thermal-loop (HTL) corrections to the
  $C\!P$-asymmetries in neutrino and, at high temperature, Higgs boson
  decays in leptogenesis. We pay special attention to the two leptonic
  quasiparticles that arise at non-zero temperature and find that
  there are four contributions to the $C\!P$-asymmetries, which
  correspond to the four combinations of the two leptonic
  quasiparticles in the loop and in the final states. In two
  additional cases, we approximate the full HTL-lepton propagator with
  a zero-temperature propagator that employs the thermal lepton mass
  $m_\ell(T)$, or the asymptotic thermal lepton mass $\sqrt{2} \,
  m_\ell(T)$. We find that the $C\!P$-asymmetries in the one-mode
  approaches differ by up to one order of magnitude from the full
  two-mode treatment in the interesting temperature regime $T \sim
  M_1$. The asymmetry in Higgs boson decays turns out to be two orders
  of magnitude larger than the asymmetry in neutrino decays in the
  zero-temperature treatment. The effect of HTL corrections on the
  final lepton asymmetry are investigated in paper II of this series.

\end{abstract}

\section{Introduction}
\label{sec:introduction}
The matter-antimatter asymmetry of the universe is usually expressed as
\begin{align}
  \label{eq:7}
  \h \equiv \left. \frac{\nb - \nbb}{\ng} \right |_0 =(6.16 \pm 0.16)
% WMAP + BAO + H_0 : 6.18 \pm 0.15
  \times 10^{-10} \, ,
\end{align}
which has been inferred from the 7-year WMAP cosmic microwave
background (CMB) anisotropy data~\cite{Komatsu:2010fb} 
% combined with the latest distance measurements from the Baryon
% Acoustic Oscillations (BAO)~\cite{Reid:2009xm} and the Hubble
% constant ($H_0$) measurement~\cite{Riess:2009pu}
, where $\nb$, $\nbb$, and $\ng$ are the number densities of baryons,
antibaryons, and photons, and the subscript 0 implies present cosmic
time.  Leptogenesis~\cite{Fukugita:1986hr} is a very attractive model
in this context since it simultaneously explains the baryon asymmetry
and the smallness of neutrino masses via the seesaw
mechanism~\cite{Minkowski:1977sc,Yanagida:1979as,GellMann:1980vs,
  Mohapatra:1980yp,Schechter:1980gr,Schechter:1981cv}. We add three
heavy right-handed neutrinos $N_i$ to the SM, which are assumed to
have rather large Majorana masses $M_i$, close to the scale of some
possibly underlying grand unified theory (GUT), $E_{\rm GUT} \sim
10^{16} \, {\rm GeV}$. In the early universe, the heavy neutrinos
decay into leptons and Higgs bosons and create a lepton asymmetry,
which is later on converted to a baryon asymmetry by the anomalous
sphaleron processes~\cite{Klinkhamer:1984di,Kuzmin:1985mm}.  The three
Sakharov conditions~\cite{Sakharov:1967dj} that are necessary for a
baryogenesis theory are fulfilled, that is lepton number $L$ and $B-L$
are violated, $C\!P$ symmetry is violated in the decays and inverse
decays and the interactions can be out of equilibrium.

Ever since the development of the theory 25 years ago, the
calculations of leptogenesis dynamics have become more refined and
many effects and scenarios that have initially been neglected have
been considered\footnote{For an excellent review of the development in
  this field, we refer to reference\cite{Davidson:2008bu}.}. Notably
the question how the hot and dense medium of SM particles influences
leptogenesis dynamics has received increasing attention over the last
years~\cite{Covi:1997dr,Giudice:2003jh,Anisimov:2010dk,Anisimov:2010gy,
  Garny:2010nz,Beneke:2010dz,Beneke:2010wd,Garbrecht:2010sz}. At high
temperature, particles show a different behaviour than in vacuum due
to their interaction with the medium: they acquire thermal masses,
modified dispersion relations and modified helicity properties. All
these properties can be summed up by viewing the particles as thermal
quasiparticles with different behaviour than their zero-temperature
counterparts, much like the large zoo of single-particle and
collective excitations that are known in high density situations in
solid-state physics. At high temperature, notably fermions can occur
in two distinct states with a positive or negative ratio of helicity
over chirality and different dispersion relations than at zero
temperature, where these dispersion relations do not break the chiral
symmetry as a zero-temperature mass does.

Thermal effects have been considered by
references~\cite{Covi:1997dr,Giudice:2003jh,Anisimov:2010dk,Anisimov:2010gy,
  Garny:2010nz,Beneke:2010dz,Beneke:2010wd,Garbrecht:2010sz}. Notably
reference~\cite{Giudice:2003jh} performs an extensive analysis of the
effects of thermal masses that arise by resumming propagators using
the hard thermal loop (HTL) resummation within thermal field theory
(TFT). However, the authors approximated the two fermionic helicity
modes with one simplified mode that behaves like a vacuum particle
with its zero-temperature mass replaced by a thermal
mass\footnote{Moreover, an incorrect thermal factor for the
  $\CP$-asymmetry was obtained, as has been pointed out in
  reference~\cite{Garny:2010nj}.}. Due to their chiral nature, there
are serious consequences to assigning a chirality breaking mass to
fermions, hence the effects of abandoning this property should be
examined. Moreover, it seems questionable to completely neglect the
negative-helicity fermionic state which, according to TFT, will be
populated at high temperature. We argue in this study that one should
include the effect of the fermionic quasiparticles in leptogenesis
calculations, since they behave differently from zero-temperature
states with thermal masses, both conceptually and regarding their
numerical influence on the $C\!P$-asymmetry. We calculate the full
hard-thermal-loop (HTL) corrections to the $C\!P$-asymmetry in
neutrino decays and, at higher temperature, Higgs boson decays, which
have four different contributions, reflecting the four possibilities
of combining the two helicity modes of the final-state lepton with the
two modes of the lepton in the loop. As a comparison, we calculate the
asymmetries for an approach where we approximate the lepton modes with
ordinary zero-temperature states and modified masses, the thermal mass
$m_\ell(T)$ and the asymptotic mass of the positive-helicity mode,
$\sqrt{2} \, m_\ell(T)$.

This paper is the first part of a two-paper series, where the second
part is concerned with solving the Boltzmann equations using
HTL-corrected rates and $C\!P$-asymmetries~\cite{Kiessig:2011ga}. The
present work deals with these corrections to the $C\!P$-asymmetries
and is structured as follows: In section~\ref{sec:prop-at-finite-t},
we briefly review the imaginary time formalism of thermal field theory
(TFT) and discuss the hard thermal loop (HTL) resummation. In
section~\ref{sec:neutrino-higgs-boson-decays}, we review our previous
calculation for neutrino decays~\cite{Kiessig:2010pr} and present a
detailed analysis of the HTL-corrected rate for Higgs boson decays at
high temperature. The $\CP$-asymmetry for the different approaches is
the main topic of section~\ref{sec:cpas}. The $\CP$-asymmetry in the
two-mode approach consists of four different contributions due to the
two possibilities for the leptons in the loops. We present some useful
rules for performing calculations with the fermionic modes and compare
the analytical expressions for the $\CP$-asymmetries in different
cases. We restrict ourselves to the hierarchical limit where the mass
of $N_1$ is much smaller than the mass of $N_2$, that is $M_2 \gg M_1$
and assume that the contribution of $N_3$ to the $C\!P$ asymmetry is
negligible. The temperature dependence of the $\CP$-asymmetry is
discussed in detail for the one-mode approach, the two-mode approach
and the vacuum case. The differences between the asymmetries and the
physical interpretation of certain features of the asymmetries are
explained in detail.  We summarise the main insights of this work in
the conclusions and give an outlook on future work and prospects.  In
appendix~\ref{sec:frequency-sums}, we derive frequency sums for the
$C\!P$-asymmetry contributions. Analytical expressions for the
$C\!P$-asymmetry are calculated in
appendix~\ref{sec:putting-all-together}, while in
appendix~\ref{cha:other-cuts} we present analytical expressions for
the $\CP$-asymmetry contributions of the two cuts through
$\{N',\ell'\}$ and $\{N',\phi'\}$\footnote{We shamelessly stole our
  notation for the cuts in the vertex contribution from reference
  \cite{Garbrecht:2010sz}.}, which we did not consider in
section~\ref{sec:cpas}, since we are working in the hierarchical
limit. We give an analytical approximation for the $C\!P$-asymmetry in
Higgs boson decays at high temperature in the one-mode approach in
appendix~\ref{sec:appr-one-mode}.

\section{Propagators at Finite Temperature}
\label{sec:prop-at-finite-t}

When going to finite temperature~\cite{LeBellac:1996}, one has to
employ ensemble weighted expectation values of operators rather than
the vacuum expectation values, so for an operator $\hat{A}$ we get
\begin{align}
  \label{eq:1}
  \langle 0 | \hat{A} | 0 \rangle \rightarrow \langle \hat{A} \rangle_\rho \equiv {\rm tr} (\rho \hat{A}) \, .
\end{align}
There are two formalisms for calculating Green's functions at finite
temperature, the imaginary time formalism and the real time
formalism. Both are equivalent and we employ the imaginary time
formalism, where the $k_0$-integration is replaced by a sum over
discrete energies, the so-called Matsubara frequencies.

\begin{figure}
  \centering
  \includegraphics[scale=0.7]{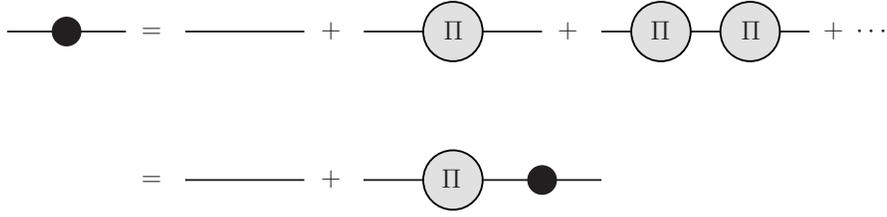}
  \caption{The resummed scalar propagator.}
  \label{fig:resummedpi}
\end{figure}

Naive perturbation theory at finite temperature can lead to serious
conceptual problems, such as infrared
divergent~\cite{Braaten:1991jj,Braaten:1991we} and gauge
dependent~\cite{Kalashnikov:1979cy, Gross:1980br} results and results
that are not complete to leading order. In order to cure these
shortcomings, the hard thermal loop (HTL) resummation technique has
been invented~\cite{Braaten:1989mz, Braaten:1990az}. One distinguishes
between hard momenta of order $T$ and soft momenta of order $gT$,
where $g$ is the coupling constant of the corresponding theory. In a
strict sense, this is only possible in the weak coupling limit where
$g \ll 1$. If all external momenta are soft, then the bare thermal
propagators have to be replaced by resummed propagators. The
self-energies that are resummed are the HTL self-energies, for which
all internal momenta are hard. For a scalar field with a
HTL-self-energy $\Pi$, the resummed effective HTL-propagator $\D^*$
follows from the Dyson-Schwinger equation in
figure~\ref{fig:resummedpi} as
\begin{align}
  \label{eq:2}
    \rmi \, \D^* & = \rmi \,\D + \rmi \, \D \left( - \rmi \, \P \right) \rmi \, \D +
  \dots \nonumber \\
  & = \frac{\rmi}{\D^{-1}-\P} = \frac{\rmi}{K^2-m_0^2-\P} \, ,
\end{align}
where $\D$ is the bare propagator, $K$ the momentum and $m_0$ the
zero-temperature mass of the scalar.
The dispersion relation for this effective excitation is given by the pole of the propagator as
\begin{align}
  \label{eq:3}
  k_0^2 = k^2+m_0^2+\Pi \, ,
\end{align}
so we get an effective mass of $m_{\rm eff}^2 = m_0^2 + m_S^2$ where
the thermal mass of the scalar is given by the self-energy, which is proportional to $gT$, $m_S^2 =
\P \propto (g T)^2$. It is possible to neglect the zero-temperature mass if $m_S \gg
m_0$.

For fermions with negligible or vanishing zero-temperature
mass, the general expression for the self-energy in the rest frame
of the thermal bath is given by~\cite{Weldon:1982bn}
\begin{align}
  \label{eq:30}
  \Sigma(P)= -a(P) \slashed{P}-b(P) \slashed{u} \, ,
\end{align}
where $u^\alpha=(1,0,0,0)$ is the four-velocity of the heat bath. The
factors $a$ and $b$ are given by
\begin{align}
  \label{eq:31}
  a(P) &= \frac{1}{4 p^2} \left[ \tr\left(\slashed{P} \Sigma\right) -
    p_0 \tr \left( \g_0 \S \right)\right] \, , \\ \nonumber
  b(P) &= \frac{1}{4 p^2} \left[P^2 \tr\left(\g_0 \Sigma\right) -
    p_0 \tr \left( \slashed{P} \S \right) \right] \, .
\end{align}
In the HTL limit, the traces are given by~\cite{LeBellac:1996}
\begin{align}
  \label{eq:32}
  T_1 \equiv \tr \left(\slashed{P} \S\right) &= 4 \, m_F^2 \, ,
  \nonumber \\
  T_2 \equiv \tr\left(\g_0\S\right) &= 2 \, m_F^2 \frac{1}{p} \ln
  \frac{p_0+p+\rmi \, \e}{p_0-p+\rmi \, \e} \, ,
\end{align}
where the effective thermal fermion mass $m_F \propto gT$ depends on the
interaction that gives rise to the fermion self-energy.
%\begin{align}
%  \label{eq:33}
%  m_F^2 =
%\begin{cases}
%  e^2 T^2/8 & \text{for QED} \\
%  g^2 T^2/16 & \text{for a Yukawa interaction } \mathcal{L}_Y = - g
%  \overline{\j} \j \f \, .
%\end{cases}
%\end{align}

The resummed fermion propagator is then written as
\begin{align}
  \label{eq:4}
    S^*(K)=\frac{1}{\slashed{K}-\Sigma_{\rm HTL}(K)}\, .
\end{align}
It is convenient to rewrite this propagator in the helicity-eigenstate
representation~\cite{Braaten:1990wp,Braaten:1992gd},
\begin{align}
  \label{eq:35}
  S^*(K)=\frac{1}{2} \Delta_+(K) (\gamma_0-\hat{\bf k} \cdot
\boldsymbol{\gamma}) +\frac{1}{2} \Delta_-(K) (\gamma_0+\hat{\bf k} \cdot
\boldsymbol{\gamma}),
\end{align}
where $\hat{\bf k}={\bf k}/k$, and
\begin{align}
\label{eq:42}
\Delta_\pm(K)=\left [ -k_0 \pm  k + \frac{m_F^2}{k} \left ( \pm 1 - 
\frac{\pm k_0 - k}{2k} \ln \frac{k_0+k}{k_0-k}  \right ) \right ]^{-1} \, .
\end{align}

\begin{figure}
  \centering
  \includegraphics[scale=0.8]{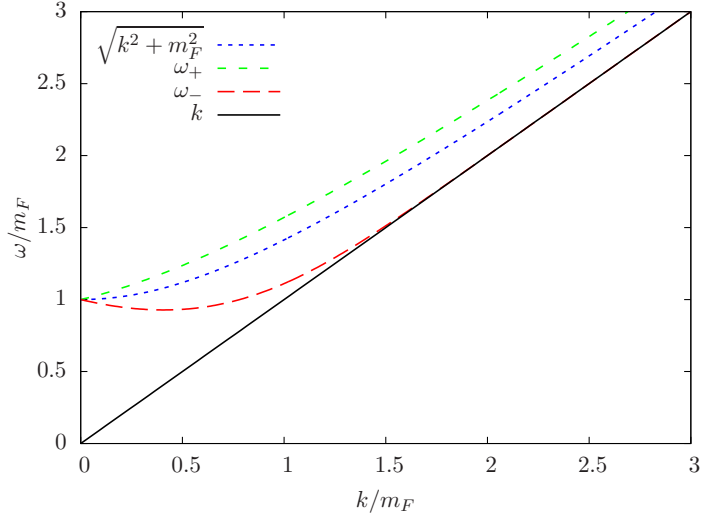}
  \caption[Fermionic dispersion relations.]{The two dispersion laws
    for fermionic excitations compared to the standard dispersion
    relation $\o^2=k^2+m_F^2$.}
  \label{fig:omega}
\end{figure}

This propagator has two poles, the zeros of the two denominators
$\D_\pm$. The poles can be seen as the dispersion relations of
single-particle excitations of the fermions that interact with the hot
plasma,
\begin{align}
  \label{eq:40}
  k_0=\o_\pm(k) \, .
\end{align}
We have presented an analytical expression for the two dispersion
relations making use of the Lambert $W$ function in
reference~\cite{Kiessig:2010pr}. The dispersion relations are shown in
figure~\ref{fig:omega}.

Note that even though the dispersion relations resemble the behaviour
of massive particles and $\o =m_F$ for zero momentum $k$, the
propagator $S^*(K)$ \eqref{eq:35} does not break chiral invariance
like a conventional mass term. Both the self energy $\S(K)$
\eqref{eq:30} and the propagator $S^*(K)$ anticommute with $\g_5$. The
Dirac spinors that are associated with the pole at $k_0=\o_+$ are
eigenstates of the operator $(\g_0-\bh{k} \cdot \gb)$ and they have a
positive ratio of helicity over chirality, $\c=+1$. The spinors
associated with $k_0=\o_-$, on the other hand, are eigenstates of
$(\g_0+\bh{k} \cdot \gb)$ and have a negative helicity-over-chirality
ratio, $\c=-1$. At zero temperature, fermions have $\c=+1$. The
introduction of a thermal bath gives rise to fermionic modes which
have $\c=-1$. These modes have been called plasminos since they are
new fermionic excitations of the plasma and have first been noted in
references~\cite{Weldon:1982bn,Klimov:1981ka}.

% \noindent \bof{Spinor eigenstates of the HTL fermion propagator}
% {\it Optional later}

We can introduce a spectral representation for the two parts of the
fermion propagator \eqref{eq:42} \cite{Pisarski:1989cs},
\begin{align}
  \label{eq:41}
  \D_\pm(K)=\int_{-\infty}^\infty \rmd
   \o \frac{\rho_\pm(\o,k)}{\o-k_0-\rmi \, \e} \, ,
\end{align}
where the spectral density $\rho_\pm(\o,k)$
\cite{Braaten:1990wp, Kapusta:1991qp} has two contributions, one from
the poles,
\begin{align}
  \label{eq:43}
  \rho^{\rm pole}_\pm(\o,k)= Z_\pm(\o,k) \, \d(\o-\o_\pm(k)) +Z_\mp(\o,k) 
  \, \d(\o+\o_\mp(k)) \, ,
\end{align}
and one discontinuous part,
\begin{align}
  \label{eq:44}
  \rho^{\rm disc}_\pm(\o,k)= \frac{\frac{1}{2}\, m_F^2(k\mp \o)}
  {\left\{k(\o\mp k)-m_F^2\left[Q_0(x) \mp Q_1(x)\right]\right\}^2 +
    \left[\frac{1}{2}\, \p\, m_F^2 (1 \mp x) \right]^2} \times
  \q(k^2-\o^2) \, ,
\end{align}
where $x=\o/k$, $\q(x)$ is the heaviside function and $Q_0$ and $Q_1$
are Legendre functions of the second kind,
\begin{align}
  \label{eq:45}
  Q_0(x)=\frac{1}{2} \ln \frac{x+1}{x-1} \, , \hspace{1cm} Q_1(x) = x \,
  Q_0(x) -1 \, .
\end{align}
The residues of the quasi-particle poles are given by
\begin{align}
  \label{eq:46}
  Z_\pm(\o,k)=\frac{\o_\pm^2(k)-k^2}{2 \, m_F^2} \, , \quad {\rm where}
  \quad Z_+ + Z_- = 1 \, .
\end{align}
One can describe the non-standard dispersion relations $\o_\pm$ by
momentum-dependent effective masses $m_\pm(k)$ which are given by
\begin{align}
  \label{eq:47}
  m_\pm(k)=\sqrt{\o_\pm^2(k)-k^2}=\sqrt{2 \, Z(\o,k)} \, m_F \, .
\end{align}
\begin{figure}
  \centering
  \includegraphics[scale=0.8]{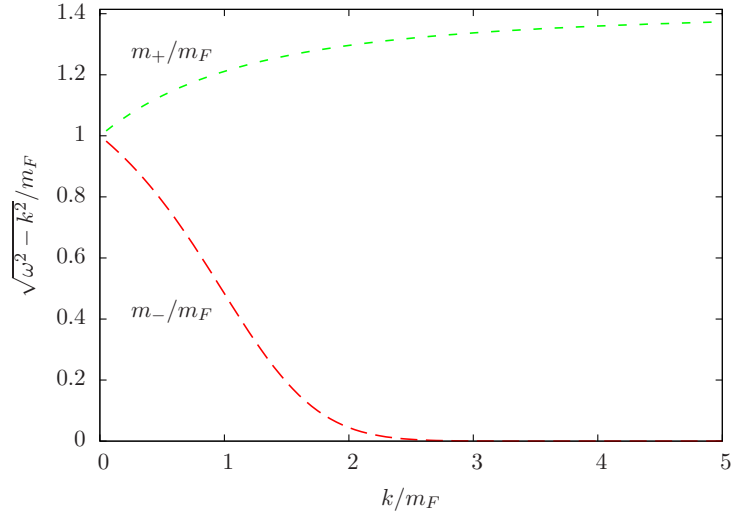}
  \caption{The momentum-dependent effective masses $m_\pm$.}
  \label{fig:mpm}
\end{figure}
These masses are shown in figure~\ref{fig:mpm}.

Considering gauge theories, one might also have to use HTL-corrected
effective vertices that are related to the propagators by Ward
identities~\cite{LeBellac:1996}. We do not consider these vertices
since we are only looking at Yukawa vertices. In the HTL framework, it
is sufficient to use bare propagators if at least one of the external
legs is hard. However, it is always possible to resum self-energies
and thus capture effects which arise from higher-order loop diagrams
and take into account the appearance of thermal masses and modified
dispersion relations in a medium. In fact, since the effective masses
we encounter do typically not satisfy the condition $m_{\rm eff} \ll
T$ but are rather in the range $m_{\rm eff}/T \sim 0.1$ -- $1$, the
effect of resummed propagators is noticeable even when some or all
external momenta are hard. In summary, we always resum the propagators
of particles that are in equilibrium with the thermal bath, which are
the Higgs bosons and the leptons in our case, in order to capture the
effects of thermal masses, modified dispersion relation and modified
helicity structures. This approach is justified a posteriori by the
sizeable corrections it reveals, similar to the treatment of meson
correlation fuctions in reference~\cite{Karsch:2000gi}.

In leptogenesis, the leptons and Higgs bosons acquire thermal masses
that have been calculated in
references~\cite{Weldon:1982bn,Klimov:1981ka,Cline:1993bd,
  Elmfors:1993re} and are given by
\begin{align}
m_\phi^2(T)&=\left (\frac{3}{16} g_2^2+ \frac{1}{16} g_Y^2 +
\frac{1}{4} y_t^2 + \frac{1}{2} \lambda \right) T^2 \, , \nonumber \\
m_\ell^2(T)&=\left (\frac{3}{32} g_2^2+ \frac{1}{32} g_Y^2 \right ) T^2.
\end{align}
The couplings denote the SU(2) coupling $g_2$, the U(1) coupling
$g_Y$, the top Yukawa coupling $y_t$ and the Higgs self-coupling
$\lambda$, where we assume a Higgs mass of about $115$ GeV. The other
Yukawa couplings can be neglected since they are much smaller than
unity and the remaining couplings are renormalised at the first
Matsubara mode, $2 \pi T$, as explained in
reference~\cite{Giudice:2003jh} and in
reference~\cite{Kajantie:1995dw} in more detail. The zero-temperature
Higgs boson mass is negligible compared to the thermal mass and the SM
fermions do not acquire a zero-temperature mass since the temperature
is above the electroweak symmetry breaking scale. The heavy neutrinos
$N_1$ do acquire a thermal mass, but since the Yukawa couplings are
much smaller than unity, this effective mass can be neglected compared
to the zero-temperature mass.

\section{Neutrino and Higgs Boson Decays}
\label{sec:neutrino-higgs-boson-decays}

In leptogenesis, we add three heavy right-handed neutrinos $N_i$ to
the SM, which are assumed to have large Majorana masses $M_i$ close to
the GUT scale, $E_{\rm GUT} \sim 10^{16} \, {\rm GeV}$. The additional
terms in the Lagrangian are
\begin{align}
  \label{eq:14}
  \cl={\rm i} \; \bar{N}_i \partial_\mu \g^\mu N_i - \l_{i \alpha}
  \bar{N_i} (\phi^a \e_{ab} \ell_\alpha^b)- \frac{1}{2} \sum_i M_i
  \bar{N}_i N_i^c + \hc \, ,
\end{align}
where the Higgs doublet $\phi$ is normalised such that its vacuum
expectation value (vev) in
\begin{equation}
\langle \phi \rangle =
\begin{pmatrix}
0 \\ v
\end{pmatrix}
\end{equation}
is $v \simeq 174 \, {\rm GeV}$ and $\l_{i \a}$ is the Yukawa coupling
connecting the Higgs doublet, the lepton doublet and the heavy
neutrino singlet. The indices $a$ and $b$ denote doublet indices and
$\e_{ab}$ is the two-dimensional total antisymmetric tensor that
ensures antisymmetric $SU(2)$-contraction.

\begin{figure}
\begin{center}
\includegraphics[scale=0.5]{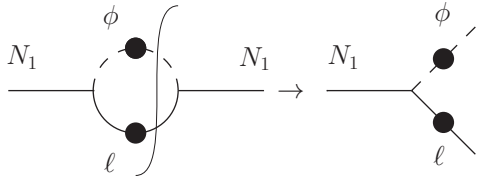}
\caption[The optical theorem in neutrino decays]{\label{cut} $N$ decay
  via the optical theorem with dressed propagators denoted by a blob.}
\end{center}
\end{figure}

We have discussed the HTL corrections to neutrino decays $N_1 \to H L$
in detail in reference~\cite{Kiessig:2010pr}. When the temperature is
so high that $m_\phi > M_1$, the neutrino decay is kinematically
forbidden in the HTL-approximation\footnote{It has been shown in
  reference~\cite{Anisimov:2010gy}, that the decay is still allowed if
  one considers the effect of collinear external momenta.}, but the
decay of Higgs bosons into neutrinos and leptons becomes
possible\footnote{The lepton decay is not possible, since $m_\phi >
  m_\ell$ for all temperatures}. The rate for the Higgs boson decays
can be calculated in the same way as the rate for the neutrino decays,
by cutting the $N_1$-self-energy with resummed lepton and Higgs boson
propagators in figure~\ref{cut}.
According to finite-temperature cutting rules
\cite{Weldon:1983jn,Kobes:1986za}, the interaction rate reads
\begin{align}
\label{eq:5}
\Gamma(P) = - \frac{1}{2 p_0} \; {\rm tr} [ (\slashed{P}+M_1) \; {\rm Im} \; 
\Sigma(p_0 + \rmi \, \e,{\bf p})].
\end{align}
At finite temperature, the self-energy for a neutrino with momentum $P$ is given by the
Matsubara sum
\begin{align}
\label{eq:6}
  \Sigma(P)=-4 \, (\l^\dagger \l)_{11} T \sum_{k_0=\rmi \, (2 n+1) \pi T} \int \frac{{\rm d}^3 k}{(2
    \pi)^3} \: P_L \: S^*(K) \: P_R \: D^*(Q),
\end{align} 
where $S^*$ and $D^*$ are the HTL-resummed lepton and Higgs boson
propagators in equations~\eqref{eq:2} and~\eqref{eq:35}, $P_L$ and
$P_R$ are the projection operators on left- and right-handed states,
$K$ is the lepton momentum and $Q=P-K$ the Higgs boson momentum. We
have summed over the two components of the lepton and Higgs doublets,
over particles and antiparticles and the three lepton flavours, so we
are looking at the processes $H \leftrightarrow N_1 L$, where the
notation $H$ and $L$ indicates that we are considering both $\phi,
\ell$ and $\barphi, \barell$.

Since the leptonic quasi-particles are the final states, we are only
interested in the pole contribution of the lepton propagator and we
get for the Matsubara sum~\cite{Kiessig:2010pr}
\begin{align}
\begin{split}
T \sum_{k_0} D^* \Delta_\pm^{\rm pole}= \frac{1}{2 \omega_q} 
& \left \{ 
\frac{\omega_\pm(k)^2-k^2}
{2 m_\ell^2} \left [ 
\frac{1+f_\phi(\o_q)-f_\ell(\o_\pm)}{p_0-\omega_\pm(k)-\omega_q}+
 \frac{f_\f(\o_q)+f_\ell(\o_\pm)}{p_0-\omega_\pm(k)+\omega_q} 
\right ]
\right. \\
& 
\label{frequencysum}
\left. +  \frac{\omega_\mp(k)^2-k^2}{2 m_\ell^2} \left [ 
\frac{f_\f(\o_q)+f_\ell(\o_\mp)}{p_0+\omega_\mp(k)-\omega_q}+
 \frac{1+f_\f(\o_q)-f_\ell(\o_\mp)}{p_0+\omega_\mp(k)+\omega_q} \right ] 
\right \} \, ,
\end{split}
\end{align} 
where $\omega_q^2 = q^2 + m_\phi^2$ is the energy of the Higgs boson,
$\o_\pm(k)$ denotes the two lepton dispersion relations, $f_\f(\o_q) =
[\exp(\o_q \b) - 1]^{-1}$ is the Bose-Einstein-distribution for the
Higgs bosons, $f_\ell(\o_\pm)=[\exp(\o_\pm \b) + 1]^{-1}$ the
Fermi-Dirac distribution for the leptons and $\b = 1/T$.

The four terms in equation~\eqref{frequencysum} correspond to the
processes with the energy relations indicated in the
denominator, i.e.~the decay $N_1 \rightarrow H
L$, the production $N_1 H \rightarrow L$, the production $N_1
L \rightarrow H$ and the production of $N_1 L H$ from the
vacuum, as well as the four inverse reactions \cite{Weldon:1983jn}. We
are only interested in the process $H \leftrightarrow N_1 L$,
where the decay and inverse decay are illustrated by the statistical
factors
\begin{align}
f_\phi+f_\ell=f_\phi (1-f_\ell)+(1+f_\phi) f_\ell \, .
\end{align}
The decay is weighted by the factor $f_\phi(1-f_\ell)$ for absorption
of a Higgs boson from the thermal bath and induced emission a lepton,
while the inverse decay is weighted by the factor $(1+f_\phi) f_\ell$
for induced emission of a Higgs boson and absorption of a lepton from
the thermal bath.  Our term reads
\begin{align}
  \label{eq:124}
  T \sum_{k_0} D^* \Delta_h \Bigg|_{H \leftrightarrow N_1 L} =  \frac{1}{2 \omega_q} \;
  \frac{\omega_{-h}^2-k^2}{2 m_\ell^2} \;
  \frac{f_\phi(\omega_q)+f_\ell(\omega_{-h})}{p_0+\omega_{-h}-\omega_q}.
\end{align}
where $h=\pm 1$ denotes the helicity-over-chirality ratio of the
final-state leptons. The angle $\eta$ between the final-state neutrino
and lepton is given by\footnote{Note that the {\it physical}
  three-momenta of the initial-state Higgs boson and the final-state
  neutrino are $- \bf q$ and $- \bf p$ since we were starting from the
  neutrino self-energy and not from the Higgs boson self-energy.}
\begin{align}
  \label{eq:113}
  \eta_h^0 = \frac{1}{2 k p} \left[ -2 p_0 \omega_h + \Sigma_\phi
  \right] \, ,
\end{align}
where 
\begin{align}
  \label{eq:122}
  \Sigma_\phi = m_\phi^2-M^2-(\omega_h^2- k^2) \, .
\end{align}
In order to clarify the momentum relations, we revert the direction of
the three-momenta ${\bf q} \to - {\bf q}$ and ${\bf p} \to -{\bf p}$
so that they correspond to the physical momenta of the incoming Higgs
boson and outgoing neutrino. The matrix element can then be derived as
\begin{align}
  \label{eq:125}
  \left| \mathcal{M}_h(P,K) \right|^2 = 4 \, (\l^\dagger \l)_{11} Z_h
  \omega_h (p_0 -h \, {\bf p} \cdot \hat{\bf k}) = 4 \, (\l^\dagger \l)_{11} P_\mu K_h^\mu \, ,
\end{align}
where we have introduced a chirally invariant four-momentum $K_h^\mu =
Z_h \o_h (1,h \, \hat{\bf k})$ for the lepton. This matrix element looks the
same as the matrix element for neutrino decays in
reference~\cite{Kiessig:2010pr}, since the momentum flip ${\bf p} \to
{\bf-p}$ compensates the helicity flip $h \to -h$ and
\begin{align}
  \label{eq:126}
  Z_h = \frac{\omega_h^2-k^2}{2 m_\ell^2} \, 
\end{align}
is the residue of the modes and the angle for the reverted physical
momenta reads
\begin{align}
  \label{eq:127}
  \eta_h^0 = \frac{1}{2 k p} \left[ 2 p_0 \omega_h - \Sigma_\phi
  \right] \, .
\end{align}

We have derived the matrix element for the Higgs boson decays starting
from the neutrino self-energy since the neutrino spinors are not
affected by thermal corrections, so we could extract the matrix
element from the expression for the neutrino interaction rate in
equation~\eqref{eq:5}. It is also possible to derive the matrix
element from the Higgs boson self energy, even though the Higgs bosons
are affected by thermal corrections, but their external states are the
same as the vacuum states since the thermal propagator in
equation~\eqref{eq:2} has the same structure as the vacuum propagator
with a different effective mass.

For leptons, the situation is different, because the structure of the
HTL-propagator in equation~\eqref{eq:35} is structurally different
from the vacuum propagator. This means that the external spinors will
have a different structure due to the modified helicity properties of
the HTL propagator. We derive important properties of the effective
lepton spinors in section~\ref{sec:defin-cpas}.

Integrating over all neutrino momenta $\bf p$, the decay densities for
neutrino and Higgs boson decay are given by
\begin{align}
  \label{eq:128}
  \gamma(N \to H L) = \int \rmd \tilde{p}_N \rmd \tilde{p}_H
  \rmd \tilde{p}_L (2 \pi)^4 \delta^4(p_N - p_L - p_H) \left|
    \mathcal{M}_h \right|^2 f_N^\rmeq (1-f_L^\rmeq)
  (1+f_H^\rmeq)
\end{align}
and
\begin{align}
  \label{eq:129}
    \gamma(H \to N L) = \int \rmd \tilde{p}_N \rmd \tilde{p}_H
  \rmd \tilde{p}_L (2 \pi)^4 \delta^4(p_N + p_L - p_H) \left|
    \mathcal{M}_h \right|^2 (1-f_N^\rmeq) (1-f_L^\rmeq)
  f_H^\rmeq \, ,
\end{align}
where $\rmd \tilde{p}=\rmd^3 p/[(2 \pi)^3 2 E]$ and the matrix element
is defined given by equation~\eqref{eq:125}.

\begin{figure}
\centering
\includegraphics[width=0.75 \textwidth]{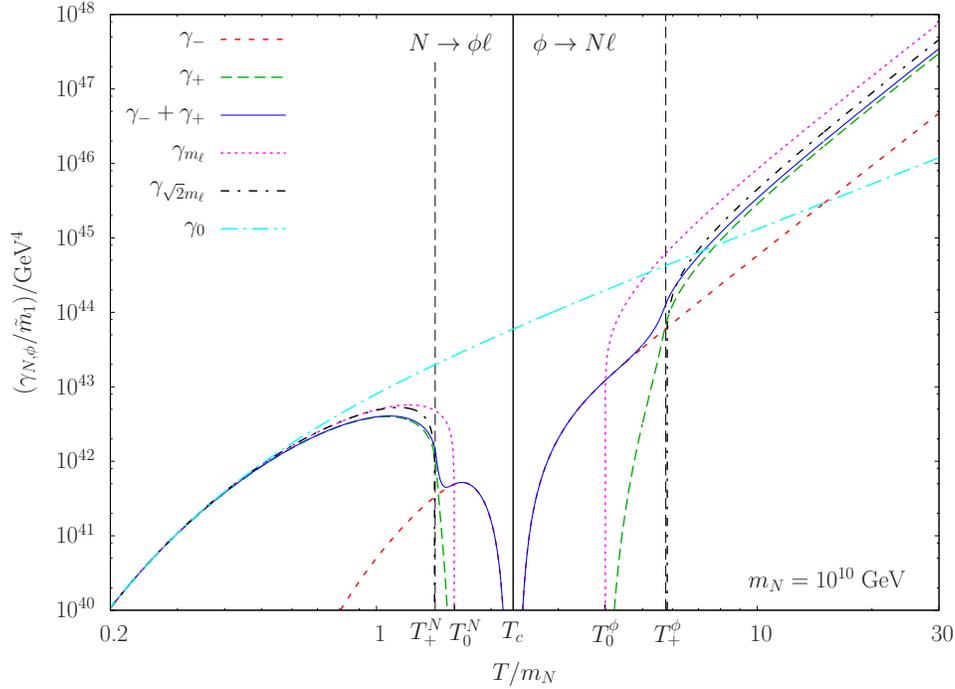}
\caption[Neutrino and Higgs boson decay densities]{The decay densities
  for the neutrino and the Higgs boson decay. We show the one-mode
  approach with the thermal mass as $\gamma_{m_{\ell}}$ and with
  the asymptotic mass as $\gamma_{\sqrt{2} \, m_{\ell}}$; Also the $T=0$ rate
  $\gamma_0$ and our two modes $\gamma_\pm$. The temperature
  thresholds are explained in the text.}
\label{comp} 
\end{figure}

In figure~\ref{comp}, we compare our consistent HTL calculation to the
one-mode approximation adopted by reference~\cite{Giudice:2003jh},
while we add quantum-statistical distribution functions to their
calculation, which equals the approach of using an approximated lepton
propagator $1/(\slashed{K}-m_\ell)$~\cite{Kiessig:2009cm}. We have
shown the decay density for the Higgs boson decays in
reference~\cite{Kiessig:2010zz}, but present a much more detailed
analysis in this work. In addition, we show the one-mode approach for
the asymptotic mass $\sqrt{2} \, m_\ell$. We evaluate the decay rates
for $M_1=10^{10}$ GeV and normalise the rates by the effective
neutrino mass $\widetilde{m}_1 = (\l^\dagger \l)_{11} v^2 / M_1$,
where $v=174$ GeV is the vacuum expectation value of the Higgs
field. This effective mass is often taken as $\widetilde{m}_1 = 0.06
\; {\rm eV}$, inspired by the mass scale of the atmospheric mass
splitting.

In the one-mode approach, the decay is forbidden when the thermal
masses of Higgs boson and lepton become larger than the neutrino mass,
$M_1<m_\ell+m_\phi$ or $M_1< \sqrt{2} \, m_\ell +m_\phi$. Considering two
modes, the kinematics exhibit a more interesting behavior. For the
plus-mode, the phase space is reduced due to the larger quasi-mass,
and at $M_1=\sqrt{2} \, m_\ell+m_\phi$, the decay is only possible into
leptons with small momenta, thus the rate drops dramatically. The
decay into the negative, quasi-massless mode is suppressed since its
residue is much smaller than the one of the plus-mode. However, the
decay is possible up to $M_1=m_\phi$. Due to the various effects, the
two-mode rate differs from the one-mode approach by more than one
order of magnitude in the interesting temperature regime of $z=T/M_1
\gtrsim 1$. The $\sqrt{2} \, m_\ell$-calculation is a better
approximation to the plus-mode, but still overestimates the rate,
which is due to the different structure of the matrix elements in the
one-mode and the two-mode approach, that is the helicity structure of
the quasiparticles. The residue also reduces the plus-rate, but the
effect is smaller since $Z_+$ is usually close to one.

At higher temperatures, when $m_\phi > M_1+m_\pm(k)$, the Higgs can
decay into neutrino and lepton modes and this process acts as a
production mechanism for neutrinos \cite{Giudice:2003jh}. The decay
$\phi \to N \ell_-$ is possible when $m_\phi > M_1$, while the decay
into $\ell_+$ is possible when $m_\phi > M_1 + m_\ell$. As for low
temperature, the rate $\gamma_+$ is unsuppressed only when $m_\phi >
M_1 + \sqrt{2} \, m_\ell$. Our decay density approaches the decay
density of reference~\cite{Giudice:2003jh} at high temperatures, but
is about a factor two below. This can be explained by the fact that
the phase space is smaller due to the larger mass of the lepton,
$m_\ell < m_h(k) < \sqrt{2} \, m_\ell$. Again, the asymptotic mass
calculation is a better approximation but still gives a larger rate
due to the matrix element and, to less extent, the residue. We see
that the decay rate rises as $\sim T^4$, instead of $\sim T^2$ as for
the vacuum rate $\gamma_0$. In the vacuum calculation, the squared
matrix element is proportional to $M_1^2$. In the finite temperature
calculation, it is proportional to $\Sigma_\phi = m_\phi^2-m_h^2(k) -
M_1^2$, so the dominant contribution is proportional to $m_\phi^2 \sim
T^2$ and the rate rises by a factor $T^2$ faster than the vacuum rate
$\gamma_0$.

Summarising, we can distinguish five different thresholds for the
thermal decay rates we discussed. Going from low temperature to high
temperature, these are given by the following conditions:
\begin{align}
T_+^N: \quad M_1&=\sqrt{2} \, m_{\ell}+m_\phi \, , \nonumber \\
T_0^N: \quad M_1&=m_{\ell}+m_\phi \, , \nonumber \\
 T_c: \quad M_1&=m_\phi \, , \nonumber \\
 T_0^\phi: \quad m_\phi&=m_{\ell}+M_1 \, , \nonumber \\
 T_+^\phi: \quad m_\phi&=\sqrt{2} \, m_{\ell}+M_1 \, .
\end{align}

\section[$C \! P$-asymmetries]{$\boldsymbol{C \! P}$-Asymmetries}
\label{sec:cpas}

\subsection{Defining $\boldsymbol{C \! P}$-asymmetries at finite temperature}
\label{sec:defin-cpas}

Let us turn to calculating the $\CP$-asymmetry in $N_1$ decays. We
denote the decaying $N_1$ by $N$ and the $N_2$ in the loop by $N'$ and
we assume that the contribution of $N_3$ to the $C\!P$ asymmetry is
negligible. At $T=0$, the $\CP$-asymmetry is defined as
\begin{align}
\label{epsilonGamma}
\epsilon_0=\frac{\Gamma(N \rightarrow \phi \ell) - 
\Gamma(N \rightarrow \bar{\phi} \bar{\ell})}
{\Gamma(N \rightarrow \phi \ell) + 
\Gamma(N \rightarrow \bar{\phi} \bar{\ell})},
\end{align}
where $\Gamma$ are the decay rates of the heavy $N$s into Higgs boson
and lepton doublet and their $\CP$-conjugated processes. At finite
temperature, we have to calculate the $\CP$-asymmetry via the
integrated decay rates,
\begin{align}
\label{epsilongamma}
\epsilon_h(T)=\frac{\gamma^{T>0}(N \rightarrow \phi \ell_h) - 
\gamma^{T>0}(N \rightarrow \bar{\phi} \bar{\ell_h})}
{\gamma^{T>0}(N \rightarrow \phi \ell_h) + 
\gamma^{T>0}(N \rightarrow \bar{\phi} \bar{\ell_h})},
\end{align}
where we define the $\CP$-asymmetry for each lepton mode, denoted by
$h$. We have
\begin{align}
\gamma^{T>0}=\int \frac{{\rm d^3}p_N}{(2 \pi)^3} f_N(p_N)
\Gamma^{T>0}(P_N^\mu) \, ,
\end{align}
where $f_N$ is the distribution function of the neutrinos and
$P_N^\mu$ the neutrino momentum. In the zero-temperature
approximation, we write
\begin{align}
\label{eq:52}
\Gamma(P^\mu)=\frac{M_1}{p_0} \Gamma_{\rm rf},
\end{align}
where $M_1$ and $p_0$ are the mass and the energy of the neutrino and
$\Gamma_{\rm rf}$ is the decay rate in the rest frame of the
neutrino. The integration over the momentum cancels out and the $\CP$
asymmetry via $\gamma$ is the same as via $\Gamma$.  At finite
temperature, however, the thermal bath breaks Lorentz invariance and
the preferred frame of reference for calculations is the rest frame of
the thermal bath. The momentum dependence of the decay rate cannot be
formulated as in equation~\eqref{eq:52} and the $\CP$-asymmetry as
defined in equation~\eqref{epsilonGamma} is momentum dependent,
therefore the definition in equation~\eqref{epsilongamma} is the
appropriate one.

The CP asymmetry in equilibrium can be written as
\begin{align}
  \epsilon_{\gamma h}^{\rm eq}(T)=\frac{ \int \frac{{\rm d^3} p}{(2
      \pi)^3} f_N (\Gamma_{Dh}-\tilde{\Gamma}_{Dh})} {\int \frac{{\rm
        d^3} p}{(2 \pi)^3} f_N (\Gamma_{D h}+\tilde{\Gamma}_{D h})},
\end{align}
where $\Gamma_{Dh}=\Gamma(N \rightarrow \ell_h \phi)$ and
$\tilde{\Gamma}_{Dh}= \Gamma(N \rightarrow \bar{\ell}_h \bar\phi)$ are the
decay rate and the $\CP$-conjugated decay rate.

The decay density is written as
\begin{align}
  \label{eq:132}
  \gamma_{Dh} = \frac{1}{2 \pi^2} \int \rmd E E p f_N^\rmeq \Gamma_{D
    h} = \frac{1}{4 (2 \pi)^3} \int \rmd E \, \rmd k \frac{k}{\o_h}
  f_N^\rmeq Z_D \left| \mathcal{M}_h \right|^2 \, .
\end{align}
where 
\begin{align}
\label{eq:167}
Z_D = (1-f_N) (1+f_\phi-f_\ell)= (1+f_\phi) (1-f_\ell)
\end{align}
is the statistical factor for the decay, with Bose-enhancement and
Fermi-blocking.  In the denominator of the $\CP$-asymmetry, it is
sufficient to take the tree-level matrix element, $\Big|
\mathcal{M}_{\rm tree} \Big|^2=\Big| \widetilde{\mathcal{M}}_{\rm
  tree} \Big|^2$. The $\CP$-asymmetry reads
\begin{align}
\epsilon_{\gamma h}(T) &= \frac{\int {\rm d}E \; {\rm d} k \;
  \frac{k}{\omega_h} \; f_N \; Z_D \;
  (|\mathcal{M}_h|^2-|\widetilde{\mathcal{M}}_h|^2)} 
{2 \; \int {\rm d}E \;
  {\rm d} k \; \frac{k}{\omega_h} \; f_N \; Z_D \; |\mathcal{M}_h|^2}
\nonumber \\
&=  \frac{1}{\gamma_h(N \to LH)} \frac{1}{4 (2 \pi)^3} \int \rmd E \rmd
k \frac{k}{\o_h} Z_D \left( \Big| \mathcal{M}_h \Big|^2 - \Big|
    \widetilde{\mathcal{M}}_h \Big|^2 \right) \, .
\end{align}

The $\CP$-asymmetry arises as the interference between tree-level and
one-loop diagrams in the decay, so we write $\mathcal{M} =
\mathcal{M}_0 + \mathcal{M}_1$, where $\mathcal{M}_0$ is the
tree-level amplitude and $\mathcal{M}_1$ the sum of all one-loop
amplitudes.  The matrix elements can be decomposed as
$\mathcal{M}_i=\lambda_i I_i$ such that the $\CP$-conjugated matrix
element is $\widetilde{\mathcal{M}_i}= \lambda_i^* I_i$. Here,
$\lambda_i$ includes the couplings and $I_i$ accounts for the
kinematics.  Thus,
\begin{align}
  |\mathcal{M}|^2-|\widetilde{\mathcal{M}}|^2= - 4 \; {\rm Im} \,
  \lambda_{CP} \; {\rm Im} \, I_{CP},
\end{align}
where $\lambda_{CP}=\lambda_0 \lambda_1^*$ and $I_{CP}= I_0 I_1^*$.

\subsection{The vertex contribution}
\label{sec:vertex-contribution}

\begin{figure}
\begin{center}
\includegraphics{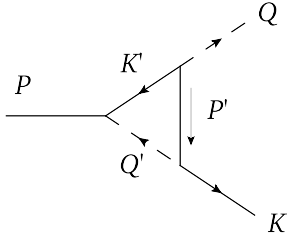}
\caption[The momentum assignments for the vertex contribution]{The
  momentum assignments for the vertex contribution to the $\CP$
  asymmetry. The solid lines without arrows are neutrinos, the ones
  with arrows the leptons and the dashed lines the Higgs bosons. All
  momenta are flowing from left to right and $P'$ as indicated.}
\label{momenta}
\end{center}
\end{figure}
Calculating the imaginary part of the kinematic term $\rmIm I_{CP}$
amounts to calculating the imaginary part of the one-loop diagram
since the tree-level diagram is real. There are two one-loop diagrams
for the neutrino decay, the vertex diagram and the self-energy
diagram. The vertex diagram is shown in figure~\ref{momenta}, along
with the momentum assignments. The coupling is
\begin{align}
  \lambda_{CP}=\lambda_0 \lambda_1^* = [(\lambda^\dagger \lambda)_{jk}]^2
  g_{SU(2)},
\end{align}
where $g_{SU(2)}=2$ denotes the sum over the Higgs and lepton
doublets, $j=1$ is the decaying neutrino family, $k=2$ is the family
of the neutrino in the loop and we have summed over all fermion spins
and the lepton families, both external and in the loop.  Moreover,
\begin{align}
\label{eq:217}
I_V= - {\rm i} \int \frac{{\rm d}^4 k'}{(2 \pi)^4} \left[ M_2
  \Delta_{N'} \Delta_{\phi'} (\overline{u}_\ell P_R u_N)
  (\overline{u}_N P_R S_{\ell'} P_L u_\ell) \right]^*,
\end{align}
where $p'$, $q'$ and $k'$ are the neutrino, Higgs boson and lepton
momentum in the loop, $\Delta_{N'}=(P'^2-M_2^2)^{-1}$ is the
denominator of the loop neutrino propagator, $\Delta_{\phi'}$
accordingly for the loop Higgs, $S_{\ell'}$ is the loop lepton
propagator and $u_N$ and $u_\ell$ are the external neutrino and lepton
spinors.

The external fermions are thermal quasiparticles and can be written as
spinors $u_\ell^\pm$~\cite{LeBellac:1996} which are eigenstates of
$(\gamma_0 \mp \hat{\bf k} \cdot \boldsymbol{\gamma})$ and have
modified dispersion relations. We have shown in
reference~\cite{Kiessig:2010pr} that
\begin{align}
\label{eq:53}
\frac{1}{2} \sum_{s}|\mathcal{M}^s_\pm(P,K)|^2=g^2 \frac{\omega_\pm^2-k^2}{2
  m_\ell^2} \omega_\pm \left (p_0 \mp p \eta_\pm \right ),
\end{align}
where $s$ denotes the spin of the neutrino. 
We can also write the matrix element as
\begin{align}
\label{eq:54}
\frac{1}{2} \sum_{s}|\mathcal{M}^s_\pm(P,K)|^2=
\frac{1}{2} \sum_{s} g^2 (\overline{u}_\ell^\pm P_R u_N^s) (\overline{u}_N^s P_L
u_\ell^\pm) \, .
\end{align}
From equations~\eqref{eq:53} and~\eqref{eq:54} we derive a rule for
multiplying the spinors of the lepton states,
\begin{align}
\label{eq:208}
u_\ell^\pm(K) \overline{u}_\ell^\pm(K) = Z_\pm
\omega_\pm (\gamma_0 \mp \hat{\bf k} \cdot
\boldsymbol{\gamma})\, ,
\end{align}
where
\begin{align}
  Z_\pm=\frac{\omega_\pm^2 - k^2}{2 m_\ell^2}
\end{align}
is the quasiparticle residuum. For the
  antiparticle spinors $v$, we replace $K$ by $-K$ and get
  \begin{align}
    \label{eq:202}
    v_\ell^\pm(K) \overline{v}_\ell^\pm(K) = - Z_\pm
\omega_\pm (\gamma_0 \pm \hat{\bf k} \cdot
\boldsymbol{\gamma}) \, .
\end{align}

The HTL lepton propagator is given in equations~\eqref{eq:35}
and~\eqref{eq:42} and the Higgs boson propagator in
equation~\eqref{eq:2}. At finite temperature, we sum over the
Matsubara modes,
\begin{align}
\int \frac{{\rm d}k'_0}{2 \pi} \rightarrow {\rm i} T \sum_{k'_0} \, ,
\end{align}
where
\begin{align}
k'_0 = (2 n + 1) \pi {\rm i} T,
\end{align}
since we are integrating over a fermion momentum.

The spin and helicity sum are evaluated as
\begin{align}
\label{eq:155}
\sum_{s,h'} (\overline{u}_\ell^h P_R u_N^s) (\overline{u}_N^s P_R S_{\ell'}^{h'} P_L
u_\ell^h) =- \sum_{h'} Z_h \omega_h M_1 \Delta_{h'} (1-h h' {\bf
  \hat{k} \cdot \hat{k'}}),
\end{align}
where $h$ and $h'$ are the ratios of helicity over chirality for the
external and the loop lepton.  The integral reads
\begin{align}
  \label{eq:131}
  I_V= - T \sum_{k_0', h'} \int \frac{{\rm d}^3 k'}{(2 \pi)^3} M_2 M_1 Z_h \o_h
  \left[
  \D_{N'} \D_{\phi'} \Delta'_{h'} \right]^* H_-^{h h'} \, ,
\end{align}
where $H_\pm = 1 \pm h h' {\bf \hat{k} \hat{k}'}$.

The frequency sum is calculated in detail in
appendix~\ref{sec:vertex-cut}. We are only interested in the
contribution from the pole part of the lepton propagator and the
explicit expression is
\begin{align}
  \label{eq:8}
\begin{split}
T \sum_{k_0'}\sum_{h'} &
\D_{N'}\D_{\phi'}\Delta_{h'}^{\rm pole}H_-=\\
= \sum_{h'}
\frac{Z_{h'}}{4 \omega_{q'} \omega_{p'}} 
& \left\{ 
\left[ \left(B_\phi^\phi-B_N^N\right)A_\ell^{\phi'}-\left( B_\phi^{\ell'}-B_N^N \right)
  A_\ell^\ell - \left(B_\phi^\phi- B_N^{\ell'}\right)A_\ell^0+
  \left(B_\phi^{\ell'}- B_N^{\ell'} \right) A_\ell^{N'}
  \right] H_- \right.\\
& \hspace{-6pt} + \left. \left[
  \left(B_\phi^{N'}-B_N^{\phi'} \right) A_\ell^{\phi'}- \left( B_\phi^0-B_N^{\phi'}
  \right) A_\ell^\ell -
  \left( B_\phi^{N'}- B_N^0 \right) A_\ell^0+ \left( B_\phi^0- B_N^0 \right) A_\ell^{N'}
  \right] H_+
\right\} \, ,
\end{split}
\end{align}
where the factors $B_{N/\phi}$ and $A_\ell$ are given by
\begin{align}
\label{eq:9}
B_{N/\phi}^\psi =& \frac{Z_{N/\phi}^\psi}{N_{N/\phi}^\psi} \, , &
A_\ell^\psi =& \frac{1}{N_\ell^\psi} \, ,
\end{align}
\begin{align}
\label{eq:10}
N_N^N & = p_0-\omega'-\omega_{q'}, & N_\ell^\ell & =
k_0-\omega_{q'}-\omega_{p'}, & N_\phi^\phi
& = q_0-\omega'-\omega_{p'}, \nonumber \\
N_N^0 & = p_0+\omega'+\omega_{q'}, & N_\ell^0 & =
k_0+\omega_{q'}+\omega_{p'}, & N_\phi^0
& = q_0+\omega'+\omega_{p'}, \nonumber \\
N_N^{\ell'} & = p_0-\omega'+\omega_{q'}, & N_\ell^{\phi'} & =
k_0-\omega_{q'}+\omega_{p'},
& N_\phi^{\ell'} & = q_0-\omega'+\omega_{p'}, \nonumber \\
N_N^{\phi'} & = p_0+\omega'-\omega_{q'}, & N_\ell^{N'} & =
k_0+\omega_{q'}-\omega_{p'}, & N_\phi^{N'} & = q_0+\omega'-\omega_{p'}
\, ,
\end{align}
\begin{align}
\label{eq:11}
Z_N^N& = 1-f_{\ell'}+f_{\phi'} \, ,
& Z_\phi^\phi &  =  1-f_{\ell'}-f_{N'} \, , \nonumber \\
Z_N^0 & = - (1 - f_{\ell'} + f_{\phi'}) \, ,
& Z_\phi^0 &  = - (1 - f_{\ell'} - f_{N'}) \, ,  \nonumber \\
Z_N^{\ell'} & = - (f_{\ell'} + f_{\phi'}) \, ,
& Z_\phi^{\ell'} & = - (f_{\ell'} - f_{N'}) \, , \nonumber \\
Z_N^{\phi'} & = f_{\ell'} + f_{\phi'} \, , & Z_\phi^{N'} & = f_{\ell'}
- f_{N'} \, .
\end{align}

\subsection{The self-energy contribution}
\label{sec:frequency-sums-self}

\begin{figure}
\begin{center}
\includegraphics{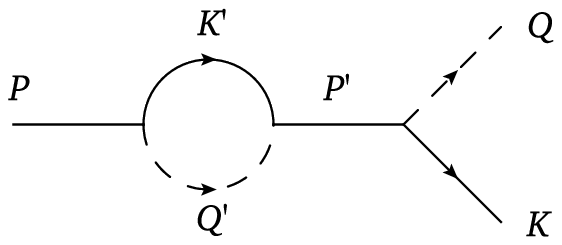}
\hspace{1.5cm}
\includegraphics{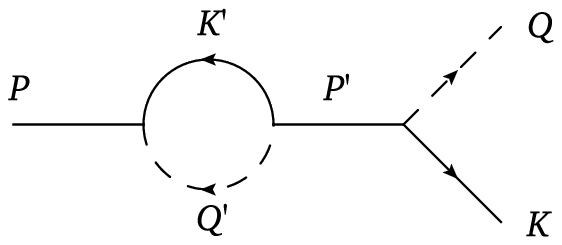}
\caption[The momentum assignments for the self-energy
contribution]{The momentum assignments for the self-energy
  contribution. The solid lines without arrows are neutrinos, the ones
  with arrows the leptons and the dashed lines the Higgs bosons.}
\label{selfenergy}
\end{center}
\end{figure}

For the self-energy contribution, the integral $I_S=I_V$ is the same as
for the vertex contribution, only the momentum relations are different
(cf.~figure~\ref{selfenergy}). The left diagram does not give a
contribution since the combination of couplings, $|(\lambda^\dagger
\lambda)_{jk}|^2$, does not have an imaginary part.

In order to carry out the Matsubara sum, we use the
Saclay-representation for the propagators. For the Higgs propagator it
is given by
\begin{align}
\Delta_{\phi}'=- \int_0^\beta {\rm d} \tau \; {\rm e}^{q_0' \tau}
\frac{1}{2 \omega_{q'}} \{ [1+f_{\phi'}(\omega_{q'})] {\rm e}^{- \omega_{q'}
  \tau} +f_{\phi'}(\omega_{q'}) {\rm e}^{\omega_{q'} \tau} \},
\label{saclayhiggs}
\end{align}
where $\omega_{q'}= \sqrt{q'^2+m_\phi^2}$ is the on-shell Higgs energy
with the thermal Higgs mass $m_\phi$ and $f_{\phi'}$ is the
Bose-Einstein distribution for the Higgs bosons with energy
$\omega_{q'}$.  For the lepton propagator the Saclay representation is
given by equations~\eqref{eq:187} and~\eqref{eq:190}. The neutrino
propagator simply reads
\begin{align}
\D_{N'}=\frac{1}{M_1^2-M_2^2},
\end{align}
since the internal neutrino momentum $P'$ is the same as the external
neutrino momentum $P$. As usual, we can write $p_0= {\rm i} \, (2m+1)
\pi T$ as Matsubara frequency and later on continue it analytically to
real values of $p_0$. In particular ${\rm e}^{p_0 \beta}=-1$.

We can calculate the frequency sum directly,
\begin{align}
T \sum_{k_0'} \rme^{q_0' \tau} \rme^{k_0' \tau'} 
= \rme^{p_0 \tau} \delta(\tau'-\tau)
\end{align}
and get
\begin{align}
T \sum_{k_0'} \D_{\phi'} \Delta'(h')=-
  \int_{-\infty}^\infty \rmd \omega' \rho'(h') \frac{1}{2 \omega_{q'}}
  \left( B_N^N - B_N^{\ell'} \right)\, .
\end{align}
Alternatively, we can use equation \eqref{eq:160} and write
\begin{align}
  \label{eq:159}
  T \sum_{k_0'} \Dt_{h',s}^{\rm pole}(k_0',\omega')
\Delta_{s_{\phi'}}(p_0-k_0',\omega_{q'}) = Z_{sh'}
\frac{s_{\phi'}}{2 \omega_{q'}} \frac{1-f_{\ell'}(s
  \omega_{sh'})+f_{\phi'}(s_{\phi'} \omega_{q'})}{p_0-s\omega_{sh'}-s_{\phi'}
  \omega_{q'}} \, .
\end{align}
Both calculations lead to
\begin{align}
T \sum_{k_0'} \sum_{h'}
  \D_{\phi'} \Delta' H_- = \sum_{h'}
  \frac{1}{2 \omega_{q'}} Z_{h'} [(B_N^N - B_N^{\ell'})
    H_- + (B_N^{\phi'} - B_N^0) H_+]\, .
\end{align}

\subsection{Imaginary parts}
\label{sec:imaginary-parts}

The terms $B_{N/\phi}^\psi$ and $A_\ell^\psi$ in the vertex
contribution in equation~\eqref{eq:102} correspond to the three
vertices where the denominator fulfills certain momentum relations
when set to zero: the $B_N$-terms correspond to the vertex with an
incoming $N_1$ and $\{\ell',\phi'\}$ in the loop, the $B_\phi$-terms
to the vertex with an outgoing $\phi$ and $\{N_2,\ell'\}$ in the loop,
and the $A_\ell$-terms to the vertex with an outgoing $\ell$ and
$\{N_2,\phi'\}$ in the loop. As an example, the term
\begin{align}
B_N^N=\frac{1-f_{\ell'}+f_{\phi'}}{p_0-\omega'-\omega_{q'}}
\end{align}
corresponds to the incoming neutrino decaying into the lepton and
Higgs boson in the loop. Thus, the terms correspond to cuttings
through the two loop lines adjacent to the vertex, however, a
correspondence with the circlings of the
RTF~\cite{LeBellac:1996,Garny:2010nj} is not obvious. Among these
cuts, only the ones which correspond to a $N_1$ or $N_2$ decaying into
a Higgs boson and a lepton are kinematically possible at the
temperatures where neutrino decay is allowed, that is where $M_1 <
m_\phi$. These terms are $B_N^N$, $A_\ell^{N'}$ and $B_\phi^{N'}$.

\subsubsection*{Regarding the $N_2$ cuts}
\label{sec:other-cuts}

\begin{figure}
  \centering
  \includegraphics[scale=0.8]{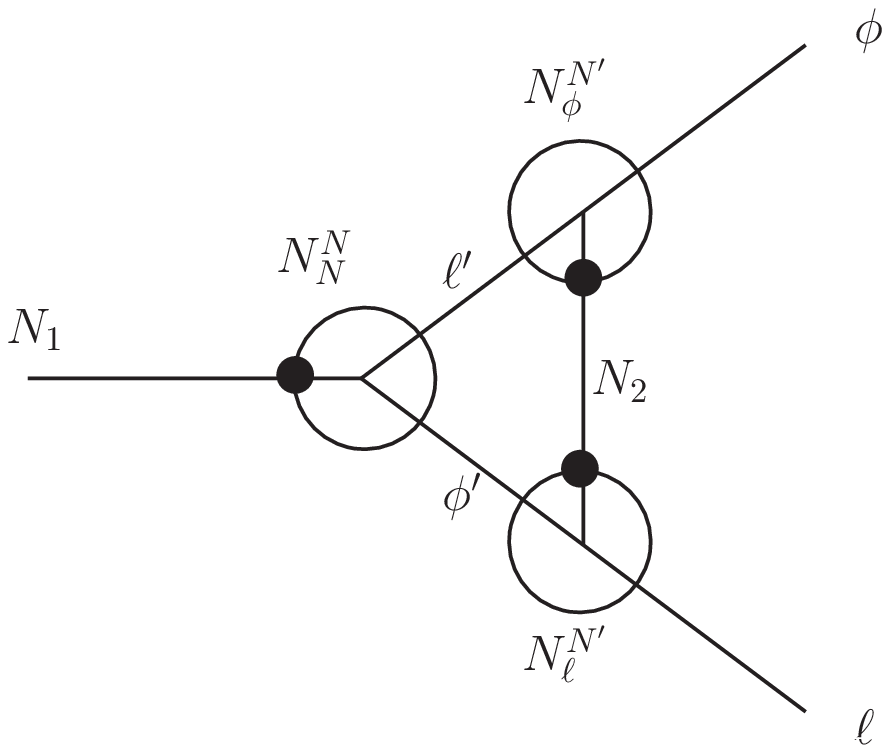}
  \caption[The cuts through the vertex contribution]{The cuts through the vertex contribution at finite
    temperature. The cuts are closed to form circles and the line
    that denotes the decaying particle in the corresponding $1 \to 2$
    process is indicated by a blob.}
  \label{fig:vertexcuts}
\end{figure}

The diagrams develop an imaginary part when one of the denominators of
the relevant terms $B_N^N$, $A_\ell^{N'}$ and $B_\phi^{N'}$
vanishes. The contributions from these denominators, $N_N^N$,
$N_\ell^{N'}$ and $N_\phi^{N'}$, correspond to the three possible cuts
shown in figure~\ref{fig:vertexcuts}. The contribution from $N_N^N$ is the
only possible cut at zero temperature. At finite temperature, the
other two cuts correspond to exchanging energy with the heat
bath. When choosing the imaginary parts corresponding to these two
cuts, the loop momentum $K'$ is of the order of $M_2$. Since we assume
a strong hierarchy $M_2 \gg M_1$, the thermal factors $f_{\phi'},
f_{\ell'}$ and $f_{N'}$ are suppressed by the large loop momentum and
the contributions become very small. In fact, they turn out to be
numerically irrelevant in the hierarchical limit.
The physical interpretation of this is as follows: Consider for
example the cut through $\{\ell',N_2\}$, which is given by a vanishing
denominator $N_\phi^{N'}$. The corresponding thermal weighting factor
is the numerator $Z_\phi^{N'}=f_{\ell'}-f_{N_2} = f_{\ell'}
(1-f_{N_2}) - (1-f_{\ell'}) f_{N_2}$. It corresponds to two processes:
absorption of a neutrino from the thermal bath and induced emission of
a lepton, or absorption of a lepton and induced emission of a
neutrino. The phase space distribution of the $N_2$s in the bath is
suppressed due to their large mass and also the distribution of
$\ell$s that have momenta large enough to fulfill momentum
conservation in the process is suppressed, so the process is
suppressed. Therefore, the thermal factors suppress the contribution
from the $N_2$-cuts. Only when we have degenerate masses $M_2 \gtrsim
M_1$, these cuts will give a contribution similar to the one from
$N_N^N$. In this case\footnote{Note that there is a mass range for
  $M_2$ where we have a contribution from the $N_2$ cuts but no
  resonant enhancement by the self-energy contribution, which becomes
  relevant when $\Delta M \equiv M_2-M_1 \ll M_1$. This mass range is
  at $\Delta M \sim M_1 \gg \G$}, the energy and temperature scales
that correspond to $N_1$ and $N_2$ processes are not clearly separated
and one has to account for the possibility of an asymmetry creation by
$N_2$ as well. Implications of these cuts were discussed in
reference~\cite{Garbrecht:2010sz}. We do not consider the influence of
this cuts, since we are working in the hierarchical limit, but we
present the analytical expression in appendix~\ref{cha:other-cuts}.

\subsubsection*{Vertex cut through $\boldsymbol{\{\ell',\phi'\}}$}
\label{sec:vertex-cut:-external-1}

The imaginary part from $N_N^N$, which implies cutting through the
lepton and Higgs boson in the loop, is the only cut that is also
possible at zero temperature and the only vertex cut that contributes
in the hierarchical limit\footnote{The corresponding $\CP$-asymmetry
  has been calculated in reference~\cite{Giudice:2003jh}, but with a
  thermal factor $1-f_{\ell'}+f_{\phi'}-2f_{\ell'} f_{\phi'}$ instead
  of the correct $1-f_{\ell'}+f_{\phi'}$. For details, see
  reference~\cite{Garny:2010nj}.}. We denote the angle between $\bf p$
and $\bf k'$ with $\eta'$,
\begin{align}
\eta'=\frac{\bf p \cdot k'}{p k'}.
\end{align}
Then
\begin{align}
{\rm Im}  \left(\int_{-1}^1 {\rmd} \eta' \frac{1}{N_N^N} \right) &= -  \pi
\int_{-1}^1 \rmd \eta' \delta(N_N^N) = -  \pi \int_{-1}^1
\frac{\omega_{q'}}{p k'} \delta(\eta' - \eta_0') \\ \nonumber
 &= -  \pi
\frac{\omega_{q'}}{p k'},
\end{align}
where the angle is
\begin{align}
\eta_0'= \frac{1}{2 p k'} \left( 2 p_0 \omega' - \Sigma_{m^2} \right)
\end{align}
and 
\begin{align}
\Sigma_{m^2}=M_1^2 + (\omega'^2 - k'^2) - m_\phi^2.
\end{align}
We get
\begin{align}
\label{imnn}
\rmIm \left( T \sum_{k_0', h'} \int \frac{\rmd^3 k'}{(2 \pi)^3}
  \D_{N'} \D_{\phi'} \Delta' H_- \right)_{N_N^N} = & \frac{1}{4 \pi^3}\rmIm \left(
  T \sum_{k_0',h'} \int_0^\infty \rmd k' k'^2 \rmd \eta' \int_0^\pi \rm
  d\phi' \D_{N'} \D_{\phi'} \Delta' H_-
\right) \nonumber \\ = & - \frac{1}{16 \pi^2} \sum_{h'} \int \rmd k'
\rmd \phi' \frac{k'}{p \omega_{p'}} Z_{h'} Z_N^N (A_\ell^\ell
-A_\ell^{\phi'}) H_-.
\end{align}
It is sufficient to perform the integration over $\phi'$ from $0$ to
$\pi$ since $\cos \phi'$ in $H_-$ is the only quantity that depends on
$\phi'$.

We note that we can write
\begin{align}
  \label{eq:162}
  A_\ell^\ell-A_\ell^{\phi'} = \frac{2
    \omega_{p'}}{(k_0-\omega_{q'})^2-\omega_{p'}^2} \equiv 2
  \omega_{p'} \Delta_{N'}^{VN} \, ,
\end{align}
where $\Delta_{N'}^{VN}$ can be viewed as the propagator of the
internal neutrino, since we can interprete the contribution we are
looking at as putting the internal Higgs boson on-shell and thus we
have $k_0-\omega_{q'} = k_0 - q_0' = p_0'$.

The analytic expression for the $C\!P$-asymmetry as defined in
equation~\eqref{epsilongamma} is worked out in
appendix~\ref{sec:all-cuts} and given by
\begin{align}
\label{eq:12}
\epsilon_h(T) = &- \frac{\rmIm\{[(\lambda^\dagger
  \lambda)_{12}]^2\}}{g_c (\lambda^\dagger \lambda)_{11}} \frac{M_1
  M_2}{4 \pi^2} \frac{\sum_{h'} \int \rmd E \rmd k \rmd k' \int_0^\pi
  \rmd \phi' k F_{N_ h}^\rmeq Z_h \frac{k'}{p \o_{p'}}Z_N^N
  Z_{h'}(A_\ell^\ell -A_\ell^{\phi'}) H_-}{\int \rmd E \rmd k k f_N
  Z_D Z_h (p_0-h p \eta)} \, ,
\end{align}
where $g_c=2$ indicates that we sum over $N \to \phi \ell$ and $N \to
\barphi \barell$ and $F_{Nh} = f_N^\rmeq (1+f_\phi^\rmeq) (1-f_{\ell
  h}^\rmeq)$ is the statistical factor for the decay.

\subsubsection*{Self-energy cut}
\label{sec:self-energy-cut}

For the self-energy diagram, only $N_N^N$ contributes. Taking $\eta'$ as
the angle between $\bf p$ and $\bf k'$, we get
\begin{align}
\rmIm \left( T \sum_{k_0', h'} \int \frac{\rmd^4 k'}{(2 \pi)^4} \sum_{h'}
  \D_{N'} \D_{\phi'} \Delta' H_- \right)_S  = & \frac{1}{4 \pi^3}\rmIm \left(
  T \sum_{k_0',h'} \int_0^\infty \rmd k' k'^2 \rmd \eta' \int_0^\pi \rm
  d\phi' \D_{N'} \D_{\phi'} \Delta' H_-
\right) \nonumber \\ = & - \frac{1}{16 \pi^2} \frac{1}{M_1^2-M_2^2} \sum_{h'} \int \rmd k'
\rmd \phi' \frac{k'}{p} Z_{h'} Z_N^N H_-.
\end{align}
Comparing this expression with the contribution from $N_N$ in
equation~\eqref{imnn}, we see that calculating the self-energy
contribution amounts to replacing $\Delta_{N'}^{VN}$ by $
\Delta_{N'}^{SN}=(M_1^2-M_2^2)^{-1}$ in the $N_N$-vertex
contribution. If $M_2 \gg M_1$, we get
\begin{align}
  \label{eq:163}
  \Delta_{N'}^{VN} \approx \Delta_{N'}^{SN} \approx -\frac{1}{M_2^2}
  \, ,
\end{align}
so the self-energy contribution is twice as large as the vertex
contribution, $\epsilon_S \approx 2 \, \epsilon_V$,
where the factor two comes from the fact that we have two
possibilities for the components of the $SU(2)$ doublets in the loop of the self-energy
diagram. This resembles the situation in vacuum.

The analytic expression for the $C\!P$-asymmetry is worked out in
appendix~\ref{sec:self-energy-cut-1} and given by
\begin{align}
\label{eq:13}
  \epsilon_h(T) = 
&- \frac{\rmIm\{[(\lambda^\dagger
  \lambda)_{12}]^2\}}{g_c (\lambda^\dagger \lambda)_{11}} \frac{M_1
  M_2}{M_1^2-M_2^2}\frac{1}{2 \pi^2}  
\frac{\sum_{h'} \int \rmd E \rmd k \rmd k'
  \int_0^\pi \rmd \phi' k F_{N_ h}^\rmeq Z_h \frac{k'}{p}Z_N^N
  Z_{h'} H_-}{\int
\rmd E \rmd k k f_N Z_D Z_h (p_0-h p \eta)} \, .
\end{align}

\subsubsection*{Symmetry under lepton-mode exchange}
\label{sec:quant-boltzm-equat}

% As derived in sections~\ref{sec:lept-antil-evol} and
% \ref{sec:lept-antil-evol-1}, the $\CP$-asymmetric quantities we need
% for the Boltzmann equations are the products $\epsilon_{\gamma
%   h}^{N/\phi} \gamma_{\epsilon h}^{N/\phi}$ in
% equations~\eqref{eq:b35} and \eqref{eq:65}. This amounts to replacing
% the factor $f_N^\rmeq (1+ f_\phi^\rmeq) (1 - f_{\ell h}^\rmeq) =
% f_\phi^\rmeq f_{\ell h}^\rmeq (1 - f_N^\rmeq)$ by $f_\phi^\rmeq
% f_{\ell h}^\rmeq (1 -2 f_N^\rmeq)$. Similarly, for the high
% temperature phase with the Higgs boson decays, we replace
% $f_\phi^\rmeq (1 - f_{\ell h}^\rmeq) (1 - f_N^\rmeq)$ by $f_\phi^\rmeq
% (1 - f_{\ell h}^\rmeq) (1 - 2 f_N^\rmeq)$.

% The relevant quantities are
% \begin{align}
%   \label{eq:104}
%   \left( \epsilon_{\gamma h}^N \gamma_{\epsilon h}^N \right)_V &=
%  - \frac{\rmIm \lambda_{CP}}{4 (2 \pi)^5} M_1 M_2 \sum_{h'} \int \rmd E
%   \rmd k \rmd k' \int_0^{\pi} \rmd \phi' k F_{N h} Z_h \frac{1}{p}
%   \frac{k'}{\omega_{p'}} Z_{h'} Z_N^N (A_\ell^\ell - A_\ell^{\phi'}) H_- \, ,
%   \nonumber \\
%   \left( \epsilon_{\gamma h}^N \gamma_{\epsilon h}^N \right)_S &=
%  - \frac{\rmIm \lambda_{CP}}{4 (2 \pi)^5} \frac{2 M_1 M_2}{M_1^2 - M_2^2}
%   \sum_{h'} \int \rmd E \rmd k \rmd k' \int_0^{\pi} \rmd \phi' k F_{N
%     h} Z_h \frac{1}{p} k' Z_{h'} Z_N^N
%   H_- \, ,
% \end{align}
% where
% \begin{align}
%   \label{eq:105}
%   F_{Nh}& = f_\phi^\rmeq
% f_{\ell h}^\rmeq (1 -2 f_N^\rmeq), \nonumber \\
%   \lambda_{CP} & =  g_{SU(2)} \left[ \left( \lambda^\dagger \lambda \right)_{12}
%     \right]^2 \, .
% \end{align}

\begin{figure}
  \centering
  \includegraphics[width=0.8 \textwidth]{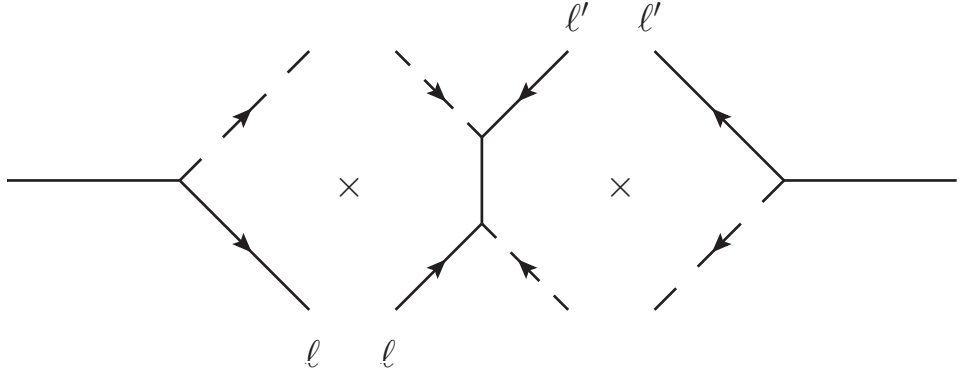}
  \caption[Factorisation of the vertex contribution]{The product of
    diagrams that corresponds to the vertex contribution of the $\CP$
    asymmetry at low temperature. It is symmetric under the exchange
    of the leptons $\ell$ and $\ell'$.}
  \label{fig:exchangeleptons}
\end{figure}

We can use equation~\eqref{eq:167} and collect all factors that depend
on $\bf k$ and $\bf k'$,
\begin{align}
  \label{eq:166}
  (1+f_\phi-f_\ell)(1+f_{\phi'}-f_{\ell'}) Z_h Z_{h'} k k' \Delta_{N'}^{VN}
  H_- \, ,
\end{align}
where we have suppressed the indices for helicity-over-chirality ratios $h$ and
$h'$. The internal neutrino momentum $\bf p = k + k' - p'$ is
symmetric under a replacement of $\bf k$ and $\bf k'$ and likewise the difference
$\omega-\omega_{q'} = \omega + \omega' - p_0$. The Higgs boson momenta
$\bf q = p - k$ and $\bf q' = p - k'$ are also exchanged when we
exchange $\bf k$ and $\bf k'$. Thus, the $\CP$-asymmetry for the
vertex contribution is symmetric under an exchange of the internal and
the external lepton. This can be understood as follows: Taking the imaginary part of $\mathcal{M}_0
\mathcal{M}_1^*$ by putting the internal lepton and Higgs boson
on-shell corresponds to calculating the product of the amplitudes of two
decays and one $\Delta L =2 $ scattering with a neutrino in the
$u$-channel, as shown in figure~\ref{fig:exchangeleptons}.
It can easily be checked that this symmetry also holds for the
self-energy diagram, where the corresponding $\Delta L=2$ scattering
has a neutrino in the $s$-channel.

\subsection{The $\boldsymbol{\CP}$-asymmetry at high temperature}
\label{sec:cp-asymmetry-at}

\begin{figure}
  \centering
    \includegraphics[scale=0.8]{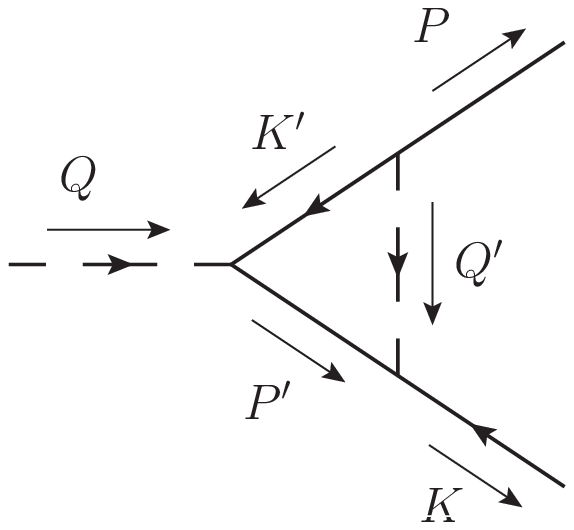} \hspace{0.5cm}
    \includegraphics[scale=0.8]{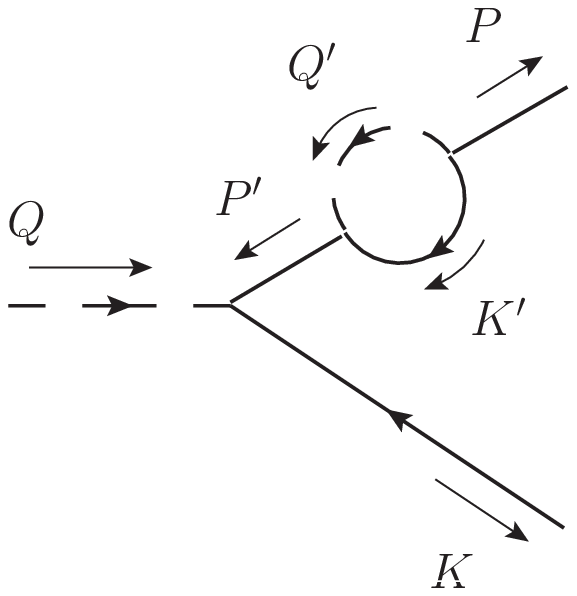}
  \caption{The vertex and the self-energy contribution for the $\phi$ decay.}
  \label{fig:phicuts}
\end{figure}

At high temperature, where we have the decays of Higgs bosons, the
$\CP$-asymmetry on amplitude level is defined as
\begin{align}
  \label{eq:101}
    \e_{h}^\phi \equiv \frac{\left| \mathcal{M}(\barphi \to N \ell_h) \right|^2 - 
\left| \mathcal{M}(\phi \to N \barell_h) \right|^2}
{\left| \mathcal{M}(\barphi \to N \ell_h) \right|^2 + 
\left| \mathcal{M}(\phi \to N \barell_h) \right|^2} \, .
\end{align}
The external momenta are now related as $q_0 = p_0+k_0$. The momentum
assignments are shown in figure~\ref{fig:phicuts}.  We take $\bf q$
and $\bf p$ as the three-momenta of the initial-state Higgs boson and
the final-state neutrino as in
section~\ref{sec:neutrino-higgs-boson-decays}, this way we can
directly use the results from the $\CP$-asymmetry in neutrino
decays. The matrix elements are the same as for the low temperature
case, so $\mathcal{M}(\phi \to N \barell_h)$ corresponds to
$\mathcal{M}(N \to \barphi \barell_h)$, just the energy relations are
different. The self-energy contribution from the external neutrino
line is the only $\CP$-asymmetric self-energy, the other self energies
do not exhibit an imaginary part in the combination of the couplings.
The couplings read
\begin{align}
  \label{eq:142}
  \rmIm \left\{ \lambda_0^\phi \lambda_1^{\phi*} \right\} = g_{SU(2)}
  \rmIm \left\{ \left[ \left( \lambda^\dagger \lambda\right)_{21}
    \right]^2 \right\} = - g_{SU(2)} \rmIm \left\{ \left[ \left(
        \lambda^\dagger \lambda\right)_{12} \right]^2 \right\} \, .
\end{align}
The integrals for the vertex and the self-energy contribution are
\begin{align}
  \label{eq:137}
    I_0^\phi I_1^{\phi*}= - T \sum_{k_0', h'} \int \frac{{\rm d}^3 k'}{(2 \pi)^3} M_2 M_1 Z_h \o_h
  \D_{N'} \D_{\phi'} \Delta'_{h'} H_-^{h h'} \, ,
\end{align}
where we remember that $\Delta_{N'}=1/(M_1^2-M_2^2)$ for the self-energy graph.

The frequency sum for the vertex diagram reads
\begin{align}
  \label{eq:143}
  \begin{split}
    T \sum_{k_0'}\sum_{h'} &
    \D_{N'}\D_{\phi'}\Delta'H_-=\\
    = \sum_{h'} \frac{Z_{h'}}{4 \omega_{q'} \omega_{p'}} & \left\{
      \left[ \left(B_\phi^\phi-B_N^N\right)A_\ell^{N'}-\left(
          B_\phi^{\ell'}-B_N^N \right) A_\ell^0 - \left(B_\phi^\phi-
          B_N^{\ell'}\right)A_\ell^\ell+ \left(B_\phi^{\ell'}-
          B_N^{\ell'} \right) A_\ell^{\phi'}
      \right] H_+ \right.\\
    & \hspace{-6pt} + \left. \left[ \left(B_\phi^{N'}-B_N^{\phi'}
        \right) A_\ell^{N'}- \left( B_\phi^0-B_N^{\phi'} \right)
        A_\ell^0 - \left( B_\phi^{N'}- B_N^0 \right) A_\ell^\ell+
        \left( B_\phi^0- B_N^0 \right) A_\ell^{\phi'} \right] H_-
    \right\} \, .
\end{split}
\end{align}

Since we have $M_2 \gg M_1$, we also have $M_2 \gg m_\phi$ in the
relevant temperature range, so the possible contributions are from
$N_N^{\phi'}$, $N_\ell^{N'}$ and $N_\phi^{N'}$\footnote{If $m_\phi \gg
  M_2$, we would have contributions from $N_\ell^{\phi'}$ and
  $N_\phi^{\phi}$ instead}. Again, the $N_2$ cuts can be neglected
because they are kinematically suppressed. When taking the
discontinuity of the diagrams, we get for the angle between $\bf p$
and $\bf k'$,
\begin{align}
  \label{eq:103}
  \eta_{\phi,0}' = \frac{1}{2 p k'} (2 p_0 \omega' -  \Sigma_\phi) \, ,
\end{align}
where
\begin{align}
  \label{eq:118}
  \Sigma_\phi = m_\phi^2 - (\omega'^2-k'^2) - M_1^2 \, ,
\end{align}
so we arrive at
\begin{align}
  \label{eq:107}
  \left( \epsilon_{\gamma h}^N \gamma_{\epsilon h}^N \right)_V &=
-  \frac{\rmIm \lambda_{CP}}{4 (2 \pi)^5} M_1 M_2 \sum_{h'} \int \rmd E
  \rmd k \rmd k' \int_0^{\pi} \rmd \phi' k F_{\phi h} Z_h \frac{1}{p}
  \frac{k'}{\omega_{p'}} Z_{h'} Z_N^{\phi'} (A_\ell^{N'} - A_\ell^0) H_- \, ,
\end{align}
where we can write
\begin{align}
  \label{eq:164}
  A_\ell^{N'} - A_\ell^{0} = \frac{2
    \omega_{p'}}{(k_0+\omega_{q'})^2-\omega_{p'}^2} = 2 \omega_{p'}
  \Delta_{N'}^{V \phi} \, .
\end{align}
Contrary to the $\CP$-asymmetry in neutrino decays, this expression
can not strictly be seen as the propagator of the neutrino since the
contribution does not correspond to a zero temperature cut but is a
pure thermal effect induced by the presence of leptons and Higgs
bosons in the thermal bath. This is illustrated by the factor
$Z_N^{\phi'}=f_{\phi'}+f_{\ell'} = f_{\phi'} (1- f_{\ell '}) + (1+
f_{\phi'}) f_{\ell'}$\footnote{Reference \cite{Giudice:2003jh} obtains
  a different factor $f_{\phi'}-f_{\ell'}-2 f_{\phi'} f_{\ell'}$ due
  to an incorrect choice of cutting rules as explained in
  reference~\cite{Garny:2010nj}.}, which describes the absorption of a
Higgs boson and the stimulated emission of a lepton and the opposite
process, the absorption of a lepton and the stimulated emission of a
Higgs boson. Compared to low temperature, we have replaced
$\Delta_{N'}^{VN} Z_N^N$ by $\Delta_{N'}^{V \phi} Z_N^{\phi'}$.

For the self-energy diagram, the frequency sum is given by
\begin{align}
  \label{eq:144}
  T \sum_{k_0'}\sum_{h'} \D_{N'}\D_{\phi'}\Delta'H_- = \Delta_{N'}
  \frac{1}{2 \omega_{q'}} \sum_{h'} Z_{h'} \left[ H_- \left(
      B_N^0-B_N^{\phi'}\right) + H_+ \left( B_N^{\ell'}- B_N^N
    \right)\right] \, ,
\end{align}
after taking the discontinuity, the $\CP$-asymmetry reads
\begin{align}
  \left( \epsilon_{\gamma h}^N \gamma_{\epsilon h}^N \right)_S &=
  - \frac{\rmIm \lambda_{CP}}{(2 \pi)^5} \frac{M_1 M_2}{M_1^2-M_2^2}
  \sum_{h'} \int \rmd E \rmd k \rmd k' \int_0^{\pi} \rmd \phi' k F_{\phi
    h} Z_h \frac{1}{p} k' Z_{h'} (-Z_N^{\phi'})
  H_- \, ,  
\end{align}
where
\begin{align}
  \label{eq:108}
  F_{\phi h}& = f_\phi^\rmeq (1- f_{\ell h}^\rmeq) (1 -f_N^\rmeq).
\end{align}
Compared to low temperature, we have replaced $Z_N^N$ by $Z_N^{\phi'}$.
The self-energy contribution is given by replacing
$\Delta_{N'}^{V \phi}$ by $\Delta_{N'}^{S \phi}=(M_1^2-M_2^2)$ in the
vertex case. For $M_2 \gg M_1$, we have 
\begin{align}
  \label{eq:165}
  \Delta_{N'}^{V \phi} \approx \Delta_{N'}^{S \phi} \approx
  -\frac{1}{M_2^2} \, ,
\end{align}
so the relation $\epsilon_S \approx 2 \epsilon_V$ also holds for the
Higgs boson decays.

Using $f_\phi (1-f_\ell) = (f_\phi + f_\ell) f_N$, the terms that
depend on the lepton momenta $\bf k$ and $\bf k'$ are
\begin{align}
  \label{eq:169}
  (f_\phi + f_\ell) (f_{\phi'}+f_{\ell'}) Z_h Z_{h'} k k' \Delta_{N'}^{S/V
    \phi} H_- \, .
\end{align}
where now $\bf p' = k+k'+p$ and $\omega+\omega_{q'} = \omega+\omega'+p_0$,
so the $\CP$-asymmetry in Higgs boson decays is symmetric under exchanging the internal and
external lepton as well.

\subsection{One-mode approach}
\label{sec:one-mode-approach}

We also calculate the $\CP$-asymmetry within the one-mode approach
where we treat the thermal mass like a kinematical mass and use lepton
propagators $(\slashed{k}-m_\ell)^{-1}$ or $(\slashed{k}-\sqrt{2}
\,m_\ell)^{-1}$ as in section~\ref{sec:neutrino-higgs-boson-decays}. The spin sum
corresponding to equation~\eqref{eq:155} then reads
\begin{align}
  \label{eq:156}
  \sum_{s,r} (\overline{u}_\ell^r P_R u_N^s) (\overline{u}_N^s P_R S_{\ell'} P_L
u_\ell^r) = 2 M_1 \Delta_{\ell'} K^\mu K_\mu' \, ,
\end{align}
where $\Delta_{\ell'}=(k_0'^2-\omega_{k'}^2)^{-1}$. In the frequency
sums in equations \eqref{eq:157} and \eqref{eq:159}, we replace
$\tilde{\Delta}_{s h'}^{\rm pole}$ by the usual decomposition
$\Delta_{s,\ell'}$, which means replacing $Z_{s h'}$ by $-s/(2
\omega')$ on the right-hand sides. One can check that in the final
expression for the $\CP$-asymmetry, this amounts to replacing the sum
of the helicity contributions
\begin{align}
  \label{eq:161}
  \sum_{h h'} Z_h
Z_{h'} (1-h h' \xi) \quad \textrm{by} \quad \frac{K^\mu
K_\mu'}{\omega_k \omega_{k'}} = 1- \frac{ k
  k'}{\omega_k \omega_{k'}} \xi \, .
\end{align}
This means that in the two mode treatment, it is forbidden for the
external and internal lepton to be scattered strictly in the same direction if
they have the same helicity or in the opposite direction if they have
opposite helicity. For the one-mode approximation this is not the case
since $\omega_k \omega_{k'}$ is always larger than $ k k'$. This result illustrates
that the leptonic quasiparticles still behave as if they are massless
in terms of the helicity structure of their interactions, while the
one-mode approach is not able to describe this behaviour.

For the $\CP$-asymmetries in the decay densities we get
\begin{align}
  \label{eq:170}
    (\Delta \gamma_m^N)_V &\equiv  \left[\gamma_m(N \to \phi \ell) -
    \gamma_m(N \to \barphi \barell) \right]_V \nonumber \\
&=
 - \frac{\rmIm \lambda_{CP}}{2 (2 \pi)^5} M_1 M_2 \sum_{h'} \int \rmd E
  \rmd k \rmd k' \int_0^{\pi} \rmd \phi' \frac{F_{N h} Z_N^N}{p} k k'
  \Delta_{N'}^{VN} \frac{K \cdot K'}{\omega_k \omega_{k'}} \, ,
  \nonumber \\
    (\Delta \gamma_m^N)_S
&=
 - \frac{\rmIm \lambda_{CP}}{(2 \pi)^5} M_1 M_2 \sum_{h'} \int \rmd E
  \rmd k \rmd k' \int_0^{\pi} \rmd \phi' \frac{F_{N h} Z_N^N}{p} k k'
  \Delta_{N'}^{SN} \frac{K \cdot K'}{\omega_k \omega_{k'}} \, ,
  \nonumber \\
    (\Delta \gamma_m^\phi)_V 
&=
 - \frac{\rmIm \lambda_{CP}}{2 (2 \pi)^5} M_1 M_2 \sum_{h'} \int \rmd E
  \rmd k \rmd k' \int_0^{\pi} \rmd \phi' \frac{F_{\phi h} Z_N^{\phi'}}{p} k k'
  \Delta_{N'}^{V \phi} \frac{K \cdot K'}{\omega_k \omega_{k'}} \, ,
  \nonumber \\
    (\Delta \gamma_m^\phi)_S
&=
 - \frac{\rmIm \lambda_{CP}}{(2 \pi)^5} M_1 M_2 \sum_{h'} \int \rmd E
  \rmd k \rmd k' \int_0^{\pi} \rmd \phi' \frac{F_{\phi h} Z_N^{\phi'}}{p} k k'
  \Delta_{N'}^{S \phi} \frac{K \cdot K'}{\omega_k \omega_{k'}} \, ,
\end{align}
where $F_{Nh} = f_N (1- f_N) (1+f_\phi -f_{\ell h})$, $F_{\phi h} = f_N (1
- f_N) (f_\phi +f_{\ell h})$.

We can examine the high temperature behaviour of the one-mode approach
by calculating the $\CP$-asymmetry in the matrix elements of a Higgs
boson at rest, where we assume that $M_1, m_\ell \ll m_\phi \ll
M_2$. The algebra is worked out in
appendix~\ref{sec:appr-one-mode}. The result reads
\begin{align}
\label{eq:15}
\epsilon_{\rm rf}^{T \gg M_1} \approx \frac{8}{g_\phi^2} \rme^{-g_\phi/2} (1
+ \rme^{-g_\phi/2}) \epsilon_0 ,
\end{align}
where $\epsilon_0$ is the $C\!P$-asymmetry in vacuum and
$g_\phi=m_\phi/T$.  Assuming that $g_\phi \ll 1$, we get
\begin{align}
\label{eq:16}
  \frac{\epsilon_{\rm rf}^{T \gg M_1}}{\epsilon_0} \approx \frac{32}{g_\phi^2}
\end{align}
Taking $g_\phi = m_\phi/T \approx 0.42$ for $T=10^{12} \, {\rm GeV}$
and using the more accurate term in equation \eqref{eq:15}, we get
$\epsilon/\epsilon_0 \approx 70$, while we get $\epsilon/\epsilon_0
\approx 90$ for equation~\eqref{eq:16}. We view this result as a
rough approximation of the value of the $\CP$-asymmetry in Higgs boson
decays at high temperature. Both our approximation and the numerical
solution of the exact expression in the next section give a factor of
100 difference to the $\CP$-asymmetry in vacuum.

\subsection{Temperature dependence of the
  $\boldsymbol{\CP}$-asymmetries}
\label{sec:results}

\begin{figure}[h]
  \centering
  \includegraphics[width=0.8 \textwidth]{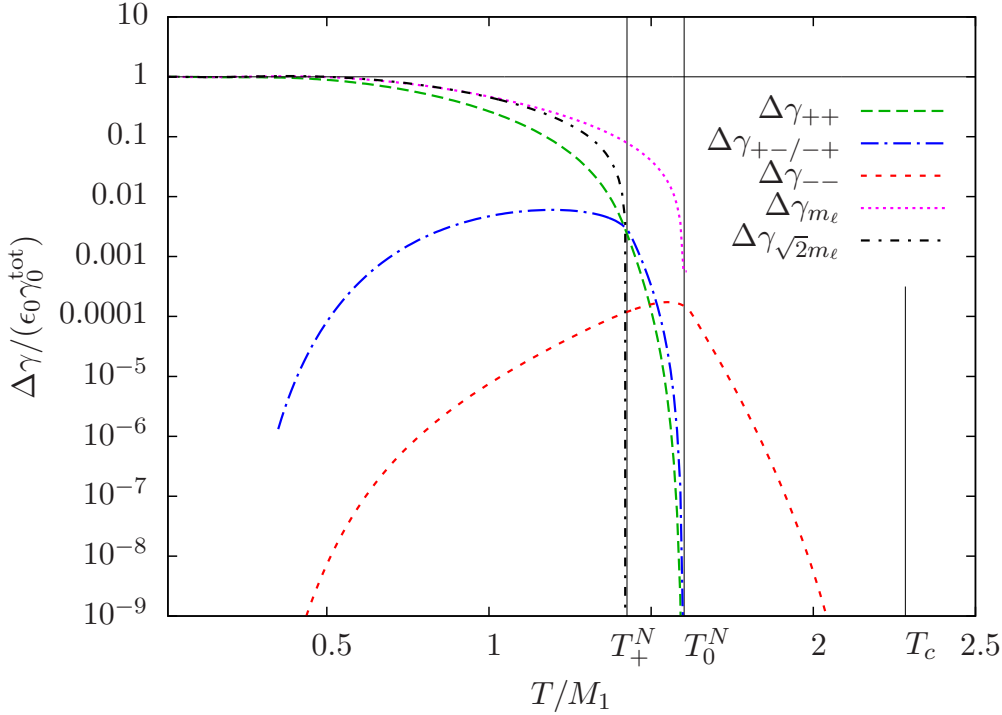}
  \caption[The $\CP$-asymmetries in $N$ decays in units of $\e_0
  \g_0$]{The $\CP$-asymmetries in neutrino decays normalised by the
    $\CP$-asymmetry in vacuum and the total decay density in vacuum,
    $\D \g/(\gamma^{\rm tot}_0 \e_0)$. We choose $M_1=10^{10} \, {\rm
      GeV}$ and $M_2 \gg M_1$. The term $\Delta \gamma_{h_1 h_2}$
    denotes the difference between the decay rate and its $\CP$
    conjugated rate, which is proportional to the
    $\CP$-asymmetry. Here, $h_1$ denotes the mode of the external
    lepton, while $h_2$ denotes the mode of the lepton in the
    loop. For example, $\Delta \gamma_{+-} = \gamma(N \to \phi
    \ell_+)- \gamma(N \to \barphi \barell_+)$, where a minus-mode
    lepton is present in the loop. $\Delta \gamma_{m_\ell}$ and
    $\Delta \gamma_{\sqrt{2} \, m_\ell}$ denote the rate differences
    for the one-mode approach with a thermal mass $m_\ell$ and an
    asymptotic thermal mass $\sqrt{2} \, m_\ell$.}
  \label{fig:cpn0}
\end{figure}

We show the temperature dependence of the $\CP$-asymmetries in neutrino
decays in the full HTL calculation and in the one-mode approach for
$m_\ell$ and $\sqrt{2} \, m_\ell$ in figure~\ref{fig:cpn0}.
We choose $M_1 = 10^{10} {\rm \, GeV}$ and normalise the asymmetries
by the product of the $\CP$-asymmetry at zero temperature and the
total decay density in vacuum, $\epsilon_0 \gamma_0^{\rm tot}$. As
discussed in sections~\ref{sec:self-energy-cut} and
\ref{sec:cp-asymmetry-at}, the vertex contribution and the self-energy
contribution have the same temperature dependence for $M_2 \gg
M_1$. Moreover, as discussed in sections~\ref{sec:quant-boltzm-equat}
and \ref{sec:cp-asymmetry-at}, the asymmetries are the same when we
exchange the internal and the external lepton, therefore the asymmetry
for a plus-mode external lepton combined with a minus-mode internal
lepton is the same as the asymmetry for a minus-mode external lepton
with a plus-mode internal lepton, in short, $\Delta \gamma_{+-} =
\Delta \gamma_{-+}$. We see that generally, the thresholds are the
ones we expect from our analysis of the decay rates in
section~\ref{sec:neutrino-higgs-boson-decays}. For the one-mode calculations we have the
expected thresholds at $T^N_0$ for $m_\ell$ and at $T^N_+$ for
$\sqrt{2} \, m_\ell$. For all asymmetries where a plus-mode lepton is
involved, that is $\Delta \gamma_{++}$, $\Delta \gamma_{+-}$ and
$\Delta \gamma_{-+}$, the phase space is reduced similar to the
$\sqrt{2} \, m_\ell$ case below $T^N_+$ and an additional reduction of
the phase space sets in between $T^N_+$ and $T^N_0$ since large
momenta $k$ or $k'$ that correspond to a large mass $m(k)$ become
kinematically forbidden. Between these two thresholds, $T_+^N$ and
$T_0^N$, the asymmetry for $\Delta \gamma_{+-/-+}$ becomes larger than
the asymmetry for $\Delta \gamma_{++}$. This effect occurs because in
the ($++$)-asymmetry the phase spaces of both the internal and the
external lepton are suppressed, while for the mixed modes, ($+-$) or
($-+$), only the phase space of one momentum is suppressed, while the
phase space of the other momentum is still large. The effect is
similar to the observation that $\gamma_-$ becomes larger than
$\gamma_+$ above $T_+^N$. Relying solely on phase-space arguments, one
would expect that $\Delta \gamma_{\sqrt{2} \, m_\ell}$ is a good
approximation for $\Delta \gamma_{++}$. The fact that $\D \g_{++}$ is
clearly smaller than $\D \g_{\sqrt{2} \, m_\ell}$ is due to two
suppressing factors: One factor is the effect of the two residues
$Z_h(k)$ and $Z_{h'}(k')$, which suppress the rate somewhat for small
momenta $k$ and $k'$. The other, more important factor is the fact
that the helicity structure and angular dependence of the integrals
are different for the ($++$)- and the $\sqrt{2} \, m_\ell$-case as
explained in section~\ref{sec:one-mode-approach}. Since neutrino
momenta are of the order $\sim M_1 \sim T$ for our temperature range,
the lepton momenta will be of the same order, that is $k > m_\ell$,
and the leptons and Higgs bosons will preferentially be scattered
forward. Thus also the angle $\xi$ between the two leptons will be
small and the factor $H_-$ defined in equation~\eqref{eq:117} is
suppressed, while the corresponding one-mode factor $1-\xi (k
k')/(\omega_k \omega_{k'})$ is larger than $H_-$ for small angles and
still finite if both leptons are scattered strictly in the same
direction, that is $\xi=1$. We have checked numerically that this is
the main reason why $\Delta \gamma_{\sqrt{2} \, m_\ell} > \Delta
\gamma_{++}$ in the range $1/2 \, M_1 \lesssim T \lesssim T_+^N$.

\begin{figure}[t]
  \centering
\includegraphics[width=0.77 \textwidth]{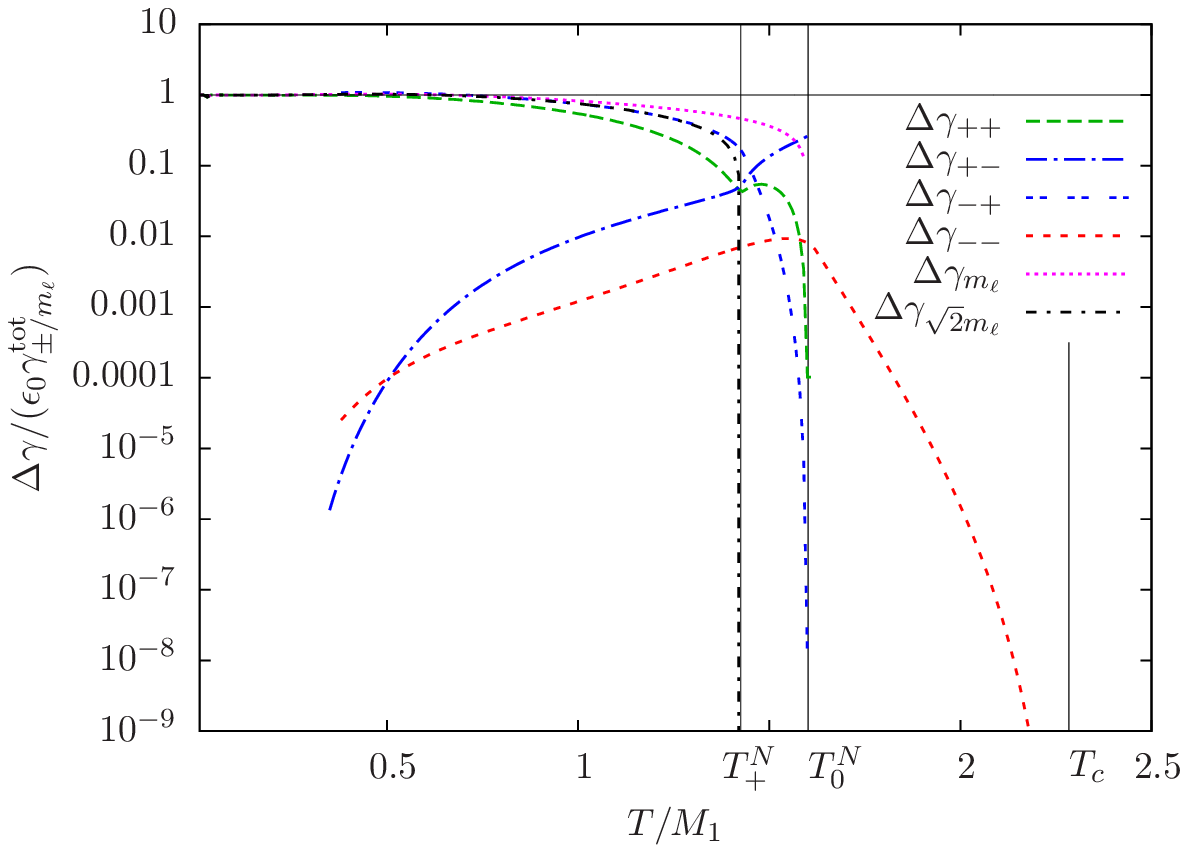}
\caption[The $\CP$-asymmetries in $N$ decays in units of $\e_0
\g^{T>0}$]{The $\CP$-asymmetries in neutrino decays normalised by the
  $\CP$-asymmetry in vacuum and the corresponding total decay density
  at finite temperature, that is $\D \g_{++}/(\gamma^{\rm tot}_+
  \e_0)$, $\D \g_{-+}/(\gamma^{\rm tot}_+ \e_0)$, $\D
  \g_{+-}/(\gamma^{\rm tot}_- \e_0)$, $\D \g_{--}/(\gamma^{\rm tot}_-
  \e_0)$, $\D \g_{m_\ell}/(\gamma^{\rm tot}_{m_\ell} \e_0)$ and $\D
  \g_{\sqrt{2} \, m_\ell}/(\gamma^{\rm tot}_{\sqrt{2} \, m_\ell}
  \e_0)$, where the $\CP$ asymmetries $\D \g$ are explained in
  figure~\ref{fig:cpn0}. We choose $M_1=10^{10} \, {\rm GeV}$ and $M_2
  \gg M_1$.}
  \label{fig:cpn1}
\end{figure}

Since the $\CP$-asymmetries follow the corresponding
finite-temperature decay rates that are shown in figure \ref{comp}, it is
very instructive to normalise them via these decay rates, that is
$\gamma_+$, $\gamma_-$, $\gamma_{m_\ell}$ and $\gamma_{\sqrt{2}\,
  m_\ell}$. This also gives a more intuitive definition of the
$\CP$-asymmetries at finite temperature. These asymmetries are shown
in figure~\ref{fig:cpn1}, normalised by the zero temperature
$\CP$-asymmetry.
Compared to the normalisation via $\gamma_0$ in figure~\ref{fig:cpn0},
we see that the ($++$)-asymmetry does not fall as steeply as the
corresponding decay rate $\gamma_+$ between $T_+^N$ and $T_0^N$, so
the ratio $\D \g_{++}/\g_+$ is dented at the threshold $T_+^N$. This
illustrates that the ($++$)-$\CP$-asymmetry shows a stronger
suppression below the threshold $T_+^N$, since it suffers from two
phase space reductions and two residues that are smaller than
one. Therefore, the $(++)$-asymmetry is not affected as strongly as
$\g_+$ by the additional suppression above $T_+^N$ when large momenta
$k$ and $k'$ are forbidden and the transition over this threshold is
smoother than for the decay rate $\g_+$. So $\gamma_+$ falls more
steeply than $\D \gamma_{++}$ above the threshold and the ratio of the
two rates has a dent at $T_+^N$. For the ($+-$)-asymmetry, this effect
is even stronger, since it is less suppressed than the
($++$)-asymmetry above $T_+^N$, so the ratio $\D \g_{+-}/(\e_0 \g_-)$
rises up to a value of $\mathcal{O}(0.1)$.
\begin{figure}
  \centering
\includegraphics[width=0.75 \textwidth]{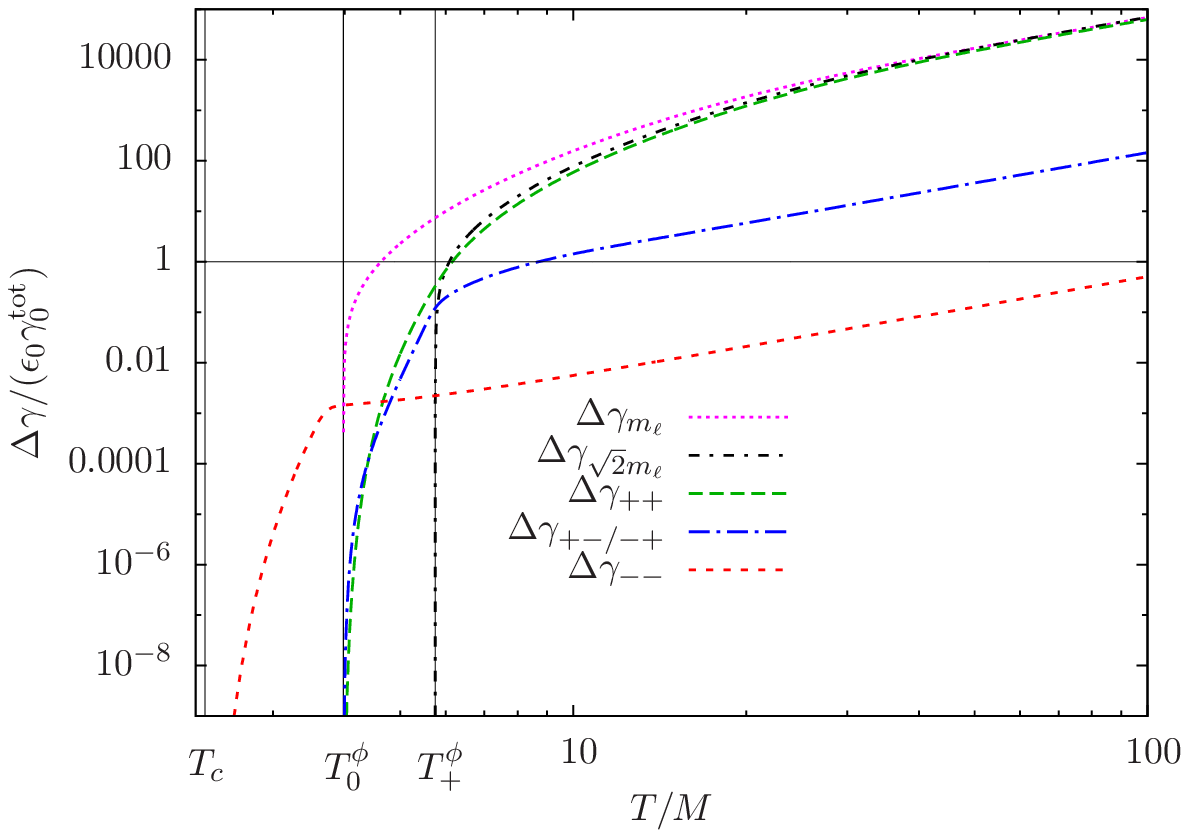}
\caption[The $\CP$-asymmetries in $\phi$ decays in units of $\e_0
\g_0$]{The $\CP$-asymmetries in Higgs boson decays normalised by the
  $\CP$-asymmetry in vacuum and the total decay density in vacuum, $\D
  \g/(\gamma^{\rm tot}_0 \e_0)$, where the asymmetries $\D \g$ are
  explained in figure~\ref{fig:cpn0}. We choose $M_1=10^{10} \, {\rm
    GeV}$ and $M_2 \gg M_1$.}
  \label{fig:cpphi0}
\end{figure}
\begin{figure}
  \centering
\includegraphics[width=0.75 \textwidth]{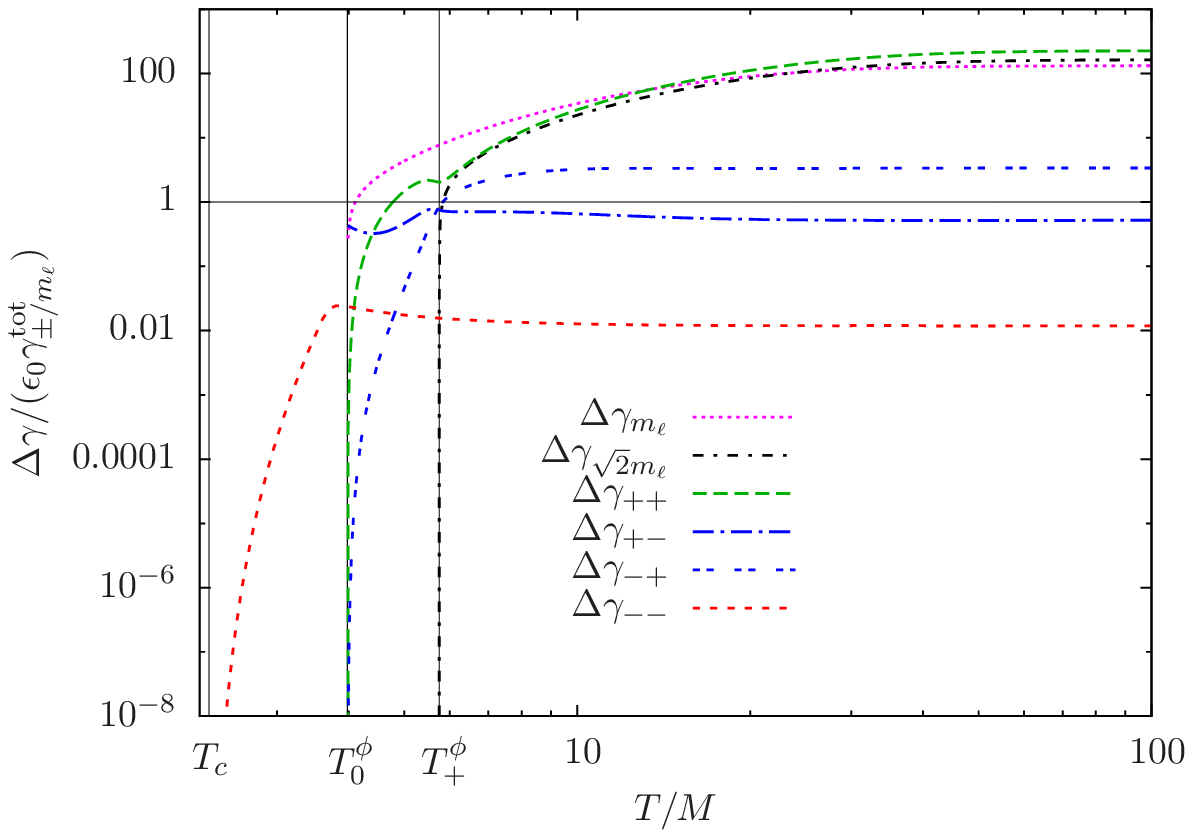}
\caption[The $\CP$-asymmetries in $\phi$ decays in units of $\e_0
\g^{T>0}$]{The $\CP$-asymmetries in Higgs boson decays normalised by
  the $\CP$-asymmetry in vacuum and the corresponding total decay
  density at finite temperature, that is $\D \g_{++}/(\gamma^{\rm
    tot}_+ \e_0)$, $\D \g_{-+}/(\gamma^{\rm tot}_+ \e_0)$, $\D
  \g_{+-}/(\gamma^{\rm tot}_- \e_0)$, $\D \g_{--}/(\gamma^{\rm tot}_-
  \e_0)$, $\D \g_{m_\ell}/(\gamma^{\rm tot}_{m_\ell} \e_0)$ and $\D
  \g_{\sqrt{2} \, m_\ell}/(\gamma^{\rm tot}_{\sqrt{2} \, m_\ell}
  \e_0)$, where the $\CP$ asymmetries $\D \g$ are explained in
  figure~\ref{fig:cpn0}. We choose $M_1=10^{10} \, {\rm GeV}$ and $M_2
  \gg M_1$.}
  \label{fig:cpphi1}
\end{figure}

The $\CP$-asymmetries in Higgs boson decays at high temperature are
shown in figure~\ref{fig:cpphi0}, normalised to $\epsilon_0
\gamma^{\rm tot}$ and we have assumed that $M_2 \gg M_1, T$.  The
behaviour is similar to the neutrino decays, where the
($--$)-asymmetry is strongly suppressed, while the ($+-$)-, the
($-+$)- and the ($++$)-asymmetries have a strict threshold at
$T_0^\phi$ and are suppressed due to the reduced phase space between
$T_0^\phi$ and $T_+^\phi$, as expected. The ($+-$)- and ($-+$)-
asymmetries are the same and are somewhat less suppressed than the
$(++)$-asymmetry between the thresholds $T_0^\phi$ and $T_+^\phi$. In
our approximation in section~\ref{sec:one-mode-approach}, we see that
the difference of the matrix elements $\Delta |\mathcal{M} |^2$ rises
as $T^2$, so $\Delta \gamma$ rises as $T^4$, which can be seen in the
plot for all finite-temperature asymmetries. The one-mode asymmetries
$\Delta \gamma_{m_\ell}$, $\Delta \gamma_{\sqrt{2} \, m_\ell}$ and the
($++$)-asymmetry are very close to each other at high temperature.

We also normalise the asymmetries to the corresponding decay rates in
figure~\ref{fig:cpphi1} and find that they all approach a constant
value at high temperature, as is expected since the decay rates also
rise as $T^4$.  The dents in the ratios $\D \g_{++}/(\e_0 \g_+)$ and
$\D \g_{+-}/(\e_0 \g_+)$ with an external plus-mode lepton are similar
to the ones for the neutrino decays and due to the very strong
suppression of $\gamma_+$ below the threshold $T_+^\phi$. The
numerically dominant asymmetries $\Delta \gamma_{m_\ell}$, $\Delta
\gamma_{\sqrt{2} \, m_\ell}$ and $\Delta \gamma_{++}$ all settle at a
rather high $\CP$-asymmetry, two orders of magnitude higher than at
zero temperature, as we expect from our approximate calculation for a
Higgs boson at rest in section~\ref{sec:one-mode-approach}. This is
partly due to a suppression of $| \mathcal{M} |^2$ which rises as
$m_\phi^2=g_\phi^2 T^2$, but mainly due to the larger difference in
matrix elements $\Delta |\mathcal{M} |^2$ for Higgs boson decays.

% 9.5.: Conclusions von Diss kopiert. Ueberarbeiten! 
% Aber erst: Paper 2 schreiben.

\section{Conclusions}
\label{sec:conclusions}

We have performed an extensive analysis of the effects of HTL
corrections on $C \! P$-asymmetries in neutrino and Higgs boson
decays. This implies capturing the effects of thermal masses, modified
dispersion relations and modified helicity structures. We put special
emphasis on the influence of the two fermionic quasiparticles, which
show a different behaviour than particles in vacuum, notably through
their dispersion relations, but also the helicity structure of their
interactions.  Our work is thus similar to the work done in
reference~\cite{Giudice:2003jh}, where the authors of the latter work
did not include the effects of fermionic quasiparticles and get a
different result for the $\CP$-asymmetries, which are crucial for the
evolution of the lepton asymmetry. We also approximate the lepton
propagators by zero temperature propagators with the zero temperature
mass replaced by the thermal lepton mass or the asymptotic mass. We
refer to these cases as one-mode approach and compare the thermal
cases to the zero-temperature case.

We have calculated HTL corrections to neutrino decays in
reference~\cite{Kiessig:2010pr} and shortly presented the Higgs decay
rate at high temperature in reference~\cite{Kiessig:2010zz}, while we
analyse this decay in detail in
section~\ref{sec:neutrino-higgs-boson-decays}. While the
zero-temperature fermion mass vanishes in the unbroken electroweak
phase, the resummation of HTL fermion self-energies results in an
effective fermion propagator that does not break chiral invariance and
is split up in two helicity modes. The external fermion states
therefore behave conceptually different from the ones with
chirality-breaking thermal masses that have been inserted in the
kinematics by hand. Moreover, one has to take care of one additional
mode, which has implications for the Boltzmann
equations~\cite{Kiessig:2011ga}.

We calculated the $\CP$-asymmetries in section~\ref{sec:cpas}. To our
knowledge, this is the first calculation of a $\CP$-asymmetry in
leptogenesis that includes HTL-corrections and takes into account
corrected thermal distributions for the particles in the
loop.\footnote{Reference\cite{Giudice:2003jh} has a calculation for
  the $\CP$-asymmetry of the neutrino and Higgs decays with HTL
  corrections in the propagators and thermal masses in the external
  states. They obtain a different combination of thermal distributions
  for the particles in the loop. The discrepancy is discussed in
  detail in reference\cite{Garny:2010nj}. The authors of the first
  reference get a factor $1-f_{\ell'}+f_{\phi'}-2 f_{\ell'} f_{\phi'}$
  for the neutrino decays and $f_{\phi'}-f_{\ell'}-2 f_{\phi'}
  f_{\ell'}$ for the Higgs boson decays due to an erroneous choice of
  cutting rules in the real time formalism. The correct calculation
  gives $1-f_{\ell'}+f_{\phi'}$ for the neutrino decays and
  $f_{\phi'}+f_{\ell'}$ for the Higgs boson decays. This discrepancy
  is also responsible for our $\CP$-asymmetry in Higgs decays being a
  factor ten larger than their result.}  We present rules for the
product of spinors that are related to the fermionic quasiparticles in
equations~\eqref{eq:202} and~\eqref{eq:208} and derive frequency sums
for the HTL fermion propagator in equation~\eqref{eq:160}. We find
four different $\CP$-asymmetries corresponding to the four different
choices of lepton modes both in the loop and as external states. We
find the $\CP$-asymmetry to be symmetric under an exchange of the
lepton mode in the loop and the external lepton mode, such that $\D
\g_{+-}= \D \g_{-+}$. At finite temperature, there are three possible
cuttings for the vertex contribution, the $\{\ell',\phi' \}$-cut that
corresponds to zero temperature and two additional cuts involving the
internal $N_2$, namely through $\{N_2,\ell'\}$ and $\{N',\phi'\}$,
which have been found by references~\cite{Giudice:2003jh,
  Beneke:2010wd, Garny:2009rv} and examined more closely in
reference~\cite{Garbrecht:2010sz}, using the real-time formalism. We
obtain the same cuts using the imaginary time formalism and
concentrate on the $\{\ell',\phi'\}$-cut, assuming the hierarchical
limit of $M_2 \gg M_1$. As expected from the zero-temperature result,
we find the vertex contribution proportional to the self-energy
contribution in this limit. Contrary to
reference~\cite{Giudice:2003jh}, we find that the $\CP$-asymmetry in
Higgs boson decays is larger than the asymmetry in neutrino decays by
about a factor of $100$ if appropriately normalised. This is due to a
suppression of the Higgs boson decay rate and a thermal enhancement of
the $\CP$-asymmetry by the distribution functions of the Higgs bosons
and leptons. We compare the $\CP$-asymmetries in the two-mode approach
to the $\CP$-asymmetries in the one-mode approach. We find that for
the two-mode approach, the helicity structure of the modes prohibits
the two leptons to be scattered strictly in the same direction while
for the one-mode approach, this direction is only mildly
suppressed. Notably this fact is responsible for suppressing the
($++$)-$\CP$-asymmetry compared to the asymmetries of the one-mode
approach, as well as the residues of the plus-modes to less extent.

Summarising, we argue that for an accurate description of medium
effects on leptogenesis, the influence of thermal quasiparticles,
notably the effects of the two fermionic modes, cannot be
neglected. The corresponding $C\!P$-asymmetries show a considerable
deviation from the one-mode asymmetries in the interesting temperature
regime $T \sim M_1$. Moreover, the presence of a new minus-mode that
essentially does not interact with the SM turns out to have
implications for the lepton asymmetry even in the strong washout
regime, as we found in an additional study~\cite{Kiessig:2011ga}.

\subsection*{Acknowledgements}

We would like to thank Mathias Garny, Georg Raffelt, Michael
A.~Schmidt and Markus Thoma for their support and comments in this
project. Thanks also to Denis Besak, Dietrich B\"odeker, Wilfried
Buchm\"uller, Valerie Domcke, Marco Drewes, Andreas Hohenegger,
Alexander Kartavtsev and Christoph Weniger for fruitful and inspiring
discussions.

\appendix

\section{Frequency Sums}
\label{sec:frequency-sums}

\subsection{Frequency sums for HTL fermion propagators}
\label{sec:frequency-sums-htl}

In order to deal with the HTL lepton propagator, we derive frequency
sums for the propagator parts $\D_\pm(K)$ of a fermion propagator. We
write the propagator in the Saclay representation as
\begin{align}
\label{eq:187}
\tilde{\Delta}_h(K)&=- \int_0^\beta \rmd \tau \, \rme^{k_0 \tau}
\Dt_h(\tau,{\bf k})\, , \nonumber \\
\Dt_h(\tau,{\bf k})&= \int_{-\infty}^\infty \rmd \omega \, \rho_h
f_F(-\omega) \rme^{-\omega \tau} \, ,
\end{align}
where $f_F$ stands for a Fermi-Dirac distribution. Since we are only
interested in the pole contribution, we write the corresponding
spectral density as
\begin{align}
  \label{eq:190}
  \rho_h^{\rm pole} &= - Z_h
  [\delta(\omega-\omega_h)+\delta(\omega+\omega_{-h})]=- \sum_s Z_{s
    h} \delta(\omega-s \omega_{s h}) \, , 
\end{align}
where $sh$ in $Z_{sh}$ and $\o_{sh}$ denotes the product of $s$ and
$h$, that is, $Z_{sh}=Z_+$ for $s=h=-1$ for example. We have for the
propagator
\begin{align}
  \label{eq:209}
  \Dt_h^{\rm pole}(\tau,{\bf k})&=- \int_{-\infty}^{\infty} \rmd \omega \, \sum_s
Z_{s h} \delta(\omega-s \omega_{s h}) f_F(-\omega) \rme^{-\omega \tau}
\nonumber \\
&= - \sum_s Z_{sh} f_F(-s \omega_{sh}) \rme^{-s \omega_{sh}\tau} 
= \sum_s \Dt_{h,s}^{\rm pole}(\tau,{\bf k}), \nonumber \\
\Dt_h^{\rm pole}(K)&= \sum_s Z_{sh} f_F(-s \omega_{sh}) \int_0^\beta
\rmd \tau \, \rme^{(k_0-s \omega_{sh})\tau} = \sum_s \Dt^{\rm
  pole}_{h,s}(K)\, , \nonumber \\
\Dt_{h,s}^{\rm pole}(K)&= Z_{sh} f_F(-s \omega_{sh}) \int_0^\beta
\rmd \tau \, \rme^{(k_0-s \omega_{sh})\tau} \nonumber \\
&= - Z_{sh} \frac{1}{k_0-s\omega_{sh}}\, ,
\end{align}
where $Z_{s h} = (\omega_{sh}^2-k^2)/(2 m_\ell^2)$ is the quasiparticle
residuum.

In dealing with frequency sums of bare thermal propagators, it is very
convenient to write
\begin{align}
\Delta_s(K)= \Delta_{-s}(-K).
\end{align}
Replacing a boson by a fermion amounts to replacing $f_B(\omega)$ by
$-f_F(\omega)$. Moreover, calculating a frequency sum of $k_0$ times
the propagators amounts to replacing $k_0$ with $s \omega$ as in
\begin{align}
T \sum_{k_0} k_0 \Delta_{s_1}(K) \Delta_{s_2}(P-K)= s_1 \omega T \sum_{k_0}
\Delta_{s_1}(K) \Delta_{s_2}(P-K),
\end{align}
where $\omega=\sqrt{k^2+m^2}$ and $m$ is the mass of the first
boson. The same holds for fermions.

It is straightforward to work out the frequency sums for the resummed
lepton propagator,
\begin{align}
\label{eq:160}
T \sum_{k_0} \Dt_{h,s_1}^{\rm pole}(k_0,\omega)
\Delta_{s_2}(p_0-k_0,\omega) = Z_{s_1 h}
\frac{s_2}{2 \omega} \frac{1-f_F(s_1
  \omega_{s_1 h})+f_B(s_2 \omega)}{p_0-s_1 \omega_{s_1 h}-s_2 \omega},
\end{align}
where the other necessary frequency sums can be derived from this by
making the appropriate substitutions.

\subsection{The frequency sum for the vertex contribution}
\label{sec:vertex-cut}

The frequency sums for the pole part of a HTL fermion propagator is
derived in appendix~\ref{sec:frequency-sums-htl}. We calculate the
frequency sum of the three propagators in the vertex loop by partial
fractioning
\begin{align}
\Dt_{s,h'}^{\rm pole} \D_{s_{\phi'}} \Dt_{s_{N'}}=C_{s\phi N'} \left[
  \frac{s_{\phi'}}{2\omega_{\phi'}}\Dt_{s,h'}^{\rm
    pole}\Dt_{s_{N'}}-\frac{s_{N'}}{2\omega_{N'}}\Dt_{s,h'}^{\rm pole}\D_{s_{\phi'}}
\right] \, .
\end{align}
We are using $\Delta_{N'}(P') = \Delta_{N'}(-P')$ and
\begin{align}
C_{s\phi N'}=\frac{1}{k_0-s_{\phi'}
  \omega_{\phi'}+s_{N'}\omega_{N'}}.
\end{align}
The frequency sum is given by
\begin{align}
\label{eq:157}
T\sum_{k_0'} \Dt_{s,h'}^{\rm pole}\D_{s_{\phi'}}\Dt_{s_{N'}}=Z_{sh'}\frac{s_{\phi'}
s_{N'}}{4\omega_{\phi'}\omega_{N'}} C_{s\phi N'}
\left[\frac{Z_{shN'}}{N_{shN'}}-
  \frac{Z_{sh\phi'}}{N_{sh\phi'}}\right],
\end{align}
where
\begin{align}
Z_{shN'}&=1-f_F(s\omega_{sh'})-f_F(s_{N'}\omega_{N'})\, , \nonumber \\
Z_{sh\phi'}&=1-f_F(s\omega_{sh'})+f(s_{\phi'}\omega_{\phi'}) \, ,
\end{align}
and
\begin{align}
N_{shN'}&=q_0-s \omega_{sh'}-s_{N'}\omega_{N'}\, , \nonumber \\
N_{sh\phi'}&=p_0-s \omega_{sh'}-s_{\phi'}\omega_{\phi'} \, .
\end{align}
Summing over all propagator parts and the helicity-over-chirality ratios, we get
\begin{align}
T \sum_{k_0'}\sum_{h'} \D_{N'}\D_{\phi'}\Delta'H_-= \sum_{h'}
\frac{Z_{h'}}{4 \omega_{q'} \omega_{p'}} \left\{E_-H_-+E_+H_+\right\},
\end{align}
where
\begin{align}
  \label{eq:117}
 H_\pm &=1 \pm h h' \xi \, , \nonumber \\
\xi &= \bf{\hat{k} \cdot \hat{k'}} \, .
\end{align}
The coefficients $E_\pm$ are given by
\begin{align}
E_-&=F^{\phi N}A_\ell^{\phi'}-F^{\ell'
  N}A_\ell^\ell -F^{\phi\ell'}A_\ell^0+F^{\ell'\ell'}A_\ell^{N'},\\
E_+&=F^{N'\phi'}A_\ell^{\phi'}-F^{0\phi'}A_\ell^\ell -F^{N'0}A_\ell^0+F^{00}A_\ell^{N'},
\end{align}
and the coefficients $F^{ij}$ read
\begin{align}
\label{eq:114}
F^{\phi N} & = B_\phi^\phi-B_N^N, & F^{N'\phi'} & = B_\phi^{N'}-B_N^{\phi'}, \nonumber \\
F^{\ell' N} & = B_\phi^{\ell'}-B_N^N, & F^{0\phi'} & = B_\phi^0-B_N^{\phi'}, \nonumber \\
F^{\phi\ell'} & = B_\phi^\phi- B_N^{\ell'}, & F^{N'0} & = B_\phi^{N'}- B_N^0, \nonumber \\
F^{\ell'\ell'} & = B_\phi^{\ell'}- B_N^{\ell'}, & F^{00} & = B_\phi^0- B_N^0.
\end{align}
The factors $B_{N/\phi}$ and $A_\ell$ are given by
\begin{align}
\label{eq:115}
B_{N/\phi}^\psi =& \frac{Z_{N/\phi}^\psi}{N_{N/\phi}^\psi} \, , &
A_\ell^\psi =& \frac{1}{N_\ell^\psi} \, ,
\end{align}
where the numerators and denominators read
\begin{align}
\label{eq:106}
  N_N^N & = p_0-\omega'-\omega_{q'}, & N_\ell^\ell & =
  k_0-\omega_{q'}-\omega_{p'}, & N_\phi^\phi 
& = q_0-\omega'-\omega_{p'}, \nonumber \\
  N_N^0 & = p_0+\omega'+\omega_{q'}, & N_\ell^0 & =
  k_0+\omega_{q'}+\omega_{p'}, & N_\phi^0 
& = q_0+\omega'+\omega_{p'}, \nonumber \\
  N_N^{\ell'} & = p_0-\omega'+\omega_{q'}, & N_\ell^{\phi'} & =
  k_0-\omega_{q'}+\omega_{p'}, 
& N_\phi^{\ell'} & = q_0-\omega'+\omega_{p'}, \nonumber \\
  N_N^{\phi'} & = p_0+\omega'-\omega_{q'}, & N_\ell^{N'} & =
  k_0+\omega_{q'}-\omega_{p'}, & N_\phi^{N'} & = 
q_0+\omega'-\omega_{p'} \, ,
\end{align}
and
\begin{align}
\label{eq:116}
Z_N^N& = 1-f_{\ell'}+f_{\phi'} \, ,
& Z_\phi^\phi &  =  1-f_{\ell'}-f_{N'} \, , \nonumber \\
Z_N^0 & = - (1 - f_{\ell'} + f_{\phi'}) \, , 
& Z_\phi^0 &  = - (1 - f_{\ell'} - f_{N'}) \, ,  \nonumber \\
Z_N^{\ell'} & = - (f_{\ell'} + f_{\phi'}) \, , 
& Z_\phi^{\ell'} & = - (f_{\ell'} - f_{N'}) \, , \nonumber \\
Z_N^{\phi'} & = f_{\ell'} + f_{\phi'} \, , 
& Z_\phi^{N'} & = f_{\ell'} - f_{N'} \, .
\end{align}
We can write
\begin{align}
\begin{split}
T \sum_{k_0'}\sum_{h'} \D_{N'}\D_{\phi'}\Delta_{h'}H_-= \sum_{h'}
\frac{Z_{h'}}{4 \omega_{q'} \omega_{p'}} & \left\{ 
\left[ F^{\phi N}A_\ell^{\phi'}-F^{\ell'
    N}A_\ell^\ell -F^{\phi\ell'}A_\ell^0+F^{\ell'\ell'}A_\ell^{N'} \right] H_- \right. \\
& \hspace{-6pt} + \left. \left[
  F^{N'\phi'}A_\ell^{\phi'}-F^{0\phi'}A_\ell^\ell -F^{N'0}A_\ell^0+F^{00}A_\ell^{N'}
  \right] H_+
\right\}
\end{split}
\end{align}
or, more explicitly,
\begin{align}
\label{eq:102}
\begin{split}
T \sum_{k_0'}\sum_{h'} &
\D_{N'}\D_{\phi'}\Delta_{h'}H_-=\\
= \sum_{h'}
\frac{Z_{h'}}{4 \omega_{q'} \omega_{p'}} 
& \left\{ 
\left[ \left(B_\phi^\phi-B_N^N\right)A_\ell^{\phi'}-\left( B_\phi^{\ell'}-B_N^N \right)
  A_\ell^\ell - \left(B_\phi^\phi- B_N^{\ell'}\right)A_\ell^0+
  \left(B_\phi^{\ell'}- B_N^{\ell'} \right) A_\ell^{N'}
  \right] H_- \right.\\
& \hspace{-6pt} + \left. \left[
  \left(B_\phi^{N'}-B_N^{\phi'} \right) A_\ell^{\phi'}- \left( B_\phi^0-B_N^{\phi'}
  \right) A_\ell^\ell -
  \left( B_\phi^{N'}- B_N^0 \right) A_\ell^0+ \left( B_\phi^0- B_N^0 \right) A_\ell^{N'}
  \right] H_+
\right\} \, .
\end{split}
\end{align}

\section{Analytic Expressions for the $\boldsymbol{C\!P}$-Asymmetries}
\label{sec:putting-all-together}

\subsection{Vertex cut through $\boldsymbol{\{\ell',\phi'\}}$}
\label{sec:all-cuts}

We simplify the analytic expression for $\epsilon_\gamma(T)$ in
equation~\eqref{epsilongamma}. For $I_V$ in equation~\eqref{eq:131}
we get
\begin{align}
\rmIm(I_V)_{N_N^N}= \frac{M_1 M_2}{16 \pi^2}
\frac{Z_h \omega}{p} \sum_{h'} \int_0^\infty \rmd k' \int_0^{\pi}
\rmd \phi' \frac{k'}{\omega_{p'}} Z_{h'} Z_N^N (A_\ell^\ell -A_\ell^{\phi'}) H_-
,
\end{align}
where it is sufficient to integrate $\phi'$ from $0$ to $\pi$. The
difference of the matrix elements reads for the vertex contribution
\begin{align}
  \label{eq:133}
  \left| \mathcal{M} (N \to \ell_h \phi) \right|^2 
 - \left|
    \mathcal{M}(N \to \barell_h \barphi) \right|^2  =& - g_{SU(2)} \rmIm
  \left\{ \left[ \left( \lambda^\dagger \lambda \right)_{12} \right]^2
  \right\} \frac{M_1 M_2}{4 \pi^2} \frac{Z_h \omega_h}{p}
  \nonumber \\
  & \times \sum_{h'}
  \int_0^\infty \rmd k' \int_0^{\pi} \rmd \phi' \frac{k'}{\omega_{p'}}
  Z_{h'} Z_N^N (A_\ell^\ell -A_\ell^{\phi'}) H_- \, .
\end{align}
Correspondingly, the difference in decay rates reads
\begin{align}
  \label{eq:134}
  \gamma(N \to \ell_h \phi) - \gamma(N \to \barell_h \barphi) = & -
  g_{SU(2)} \rmIm \left\{ \left[ \left( \lambda^\dagger \lambda
      \right)_{12} \right]^2
  \right\} \frac{M_1 M_2}{4 (2 \pi)^5} \nonumber \\
  & \times \sum_{h'} \int \rmd E \rmd k \rmd k' \int_0^\pi \rmd \phi'
  k F_{N h} Z_h \frac{k'}{p \o_{p'}}Z_N^N Z_{h'}(A_\ell^\ell
  -A_\ell^{\phi'}) H_- \, ,
\end{align}
where $F_{Nh} = f_N^\rmeq (1+f_\phi^\rmeq) (1-f_{\ell h}^\rmeq)$ is
the statistical factor for the decay.

We know from section~\ref{sec:neutrino-higgs-boson-decays} that
\begin{align}
\sum_{s} \left| \mathcal{M}^s_h(N \to LH) \right|^2 = g_{SU(2)} g_c (\lambda^\dagger \lambda)_{11}
Z_h \omega (p_0 - h p \eta) \, ,
\end{align}
where $g_c=2$ indicates that we sum over $N \to \phi \ell$ and $N \to
\barphi \barell$. Thus
\begin{align}
\Gamma(N \to L_h H) = g_{SU(2)} g_c \frac{(\lambda^\dagger \lambda)_{11}}{16 \pi p p_0} \int
\rmd k k Z_D Z_h (p_0 - h p \eta)
\end{align}
and 
\begin{align}
\gamma(N \to L_h H) =  g_{SU(2)} g_c \frac{(\lambda^\dagger
  \lambda)_{11}}{4 (2 \pi)^3} \int
\rmd E \rmd k k f_N Z_D Z_h (p_0-h p \eta) \, ,
\end{align}
where we have summed over the neutrino degrees of freedom.

We arrive at
\begin{align}
\label{eq:174}
\epsilon_h(T) = &  - g_{SU(2)} \frac{\rmIm
  \{[( \lambda^\dagger \lambda)_{12}]^2\}}{\gamma(N \to L_h N)} 
\frac{M_1 M_2}{4 (2 \pi)^5} \sum_{h'} \int \rmd E \rmd k \rmd k'
  \int_0^\pi \rmd \phi' k F_{N_ h}^\rmeq Z_h \frac{k'}{p \o_{p'}}Z_N^N
  Z_{h'}(A_\ell^\ell -A_\ell^{\phi'}) H_- \nonumber \\
=&- \frac{\rmIm\{[(\lambda^\dagger
  \lambda)_{12}]^2\}}{g_c (\lambda^\dagger \lambda)_{11}} \frac{M_1
  M_2}{4 \pi^2}  
\frac{\sum_{h'} \int \rmd E \rmd k \rmd k'
  \int_0^\pi \rmd \phi' k F_{N_ h}^\rmeq Z_h \frac{k'}{p \o_{p'}}Z_N^N
  Z_{h'}(A_\ell^\ell -A_\ell^{\phi'}) H_-}{\int
\rmd E \rmd k k f_N Z_D Z_h (p_0-h p \eta)} \, .
\end{align}

\subsection{Self-energy cut}
\label{sec:self-energy-cut-1}

For the self-energy contribution, we get
\begin{align} 
\rmIm(I_S)_{N_N}=  \frac{M_1 M_2}{M_1^2-M_2^2} \frac{1}{4 \pi^2}
\frac{Z_h \omega}{p} \sum_{h h'} \int_0^\infty \rmd k' \int_0^{\pi}
\rmd \phi' k' Z_{h'} Z_N^N H_- \, .
\end{align}
The difference in decay rates reads
\begin{align}
  \label{eq:140}
     \gamma(N \to \ell_h \phi) - \gamma(N \to \barell_h \barphi) = 
& - g_{SU(2)} \rmIm
  \left\{ \left[ \left( \lambda^\dagger \lambda \right)_{12} \right]^2
  \right\} \frac{M_1 M_2}{(M_1^2-M_2^2)}\frac{1}{(2 \pi)^5} \nonumber \\
 & \times \sum_{h'} \int \rmd E \rmd k \rmd k'
  \int_0^\pi \rmd \phi' k F_{N_ h}^\rmeq Z_h \frac{k'}{p}Z_N^N
  Z_{h'} H_- \, .
\end{align}
The $\CP$-asymmetry reads
\begin{align}
  \label{eq:141}
  \epsilon_h(T) = &  - g_{SU(2)} \frac{\rmIm
  \{[( \lambda^\dagger \lambda)_{12}]^2\}}{\gamma(N \to L_h N)} 
\frac{M_1 M_2}{M_1^2-M_2^2}\frac{1}{(2 \pi)^5} \sum_{h'} \int \rmd E \rmd k \rmd k'
  \int_0^\pi \rmd \phi' k F_{N_ h}^\rmeq Z_h \frac{k'}{p}Z_N^N
  Z_{h'} H_- \nonumber \\
=&- \frac{\rmIm\{[(\lambda^\dagger
  \lambda)_{12}]^2\}}{g_c (\lambda^\dagger \lambda)_{11}} \frac{M_1
  M_2}{M_1^2-M_2^2}\frac{1}{2 \pi^2}  
\frac{\sum_{h'} \int \rmd E \rmd k \rmd k'
  \int_0^\pi \rmd \phi' k F_{N_ h}^\rmeq Z_h \frac{k'}{p}Z_N^N
  Z_{h'} H_-}{\int
\rmd E \rmd k k f_N Z_D Z_h (p_0-h p \eta)} \, ,
\end{align}

\section{The Other Cuts}
\label{cha:other-cuts}

\subsection{Imaginary Parts}
\label{sec:imaginary-parts-1}

\subsubsection*{Vertex cut through $\boldsymbol{\{N_2,\phi'\}}$}
\label{sec:vertex-cut:-external-2}

We use the conventions for the vertex contribution in $N$-decays in
section~\ref{sec:cpas}. For $N_\ell^{N'}$, we shift integration
variables to $\rmd^3 q'$ after carrying out the Matsubara sum over
$k_0'$. We consider the angle
\begin{align}
\eta_{q'}= \frac{\bf k \cdot q'}{k q'}
\end{align}
between $\bf k$ and $\bf q'$ and write
\begin{align}
\rmIm  \left( \int_{-1}^1 \rmd \eta_{q'} \frac{1}{N_\ell^{N'}} \right) = -\pi
\frac{\omega_{p'}}{k q'},
\end{align}
where the angle is
\begin{align}
\eta_{k q', 0}= \frac{1}{2 k q'} \left(- 2 \omega \omega_{q'} +
\Sigma_k \right),
\end{align}
and
\begin{align}
\Sigma_k=M_2^2-(\omega^2-k^2)-m_\phi^2 \, .
\end{align}
The imaginary part reads
\begin{align}
\label{imnn'}
 \rmIm  \Big( T \sum_{k_0', h'}\int \frac{\rmd^3 k'}{(2 \pi)^3}
 \D_{N'} \D_{\phi'} & \D_{h'} H_-
  \Big)_{N_\ell^{N'}} =  \frac{1}{4 \pi^3}\rmIm \left(
  T \sum_{k_0',h'} \int_0^\infty \rmd q' q'^2 \rmd \eta_{q'}
  \int_0^\pi \rmd \phi_{q'} \D_{N'} \D_{\phi'} \D_{h'} H_-
\right) \nonumber \\
 =& - \frac{1}{16 \pi^2} \sum_{h'} \int \rmd q'
\rmd \phi' \frac{q'}{k \omega_{q'}} Z_{h'} 
\left[ \left( B_\phi^{\ell'} - B_N^{\ell'} \right) H_- + \left(
    B_\phi^0  - B_N^0 \right) H_+ \right] \, .
\end{align}

\subsubsection*{Vertex cut through $\boldsymbol{\{N_2,\ell'\}}$}
\label{sec:vertex-cut:-external-3}

For the $N_\phi^{N'}$-term, we integrate
over $k'$, and choose the polar angle between $\bf q$ and $\bf k'$,
\begin{align}
\eta_{qk'}=\frac{\bf q \cdot k'}{q k'} \, .
\end{align}
We write
\begin{align}
{\rm Im} \left( \int_{-1}^1 \rmd \eta_{qk'} \frac{1}{N_{N'}}\right)=- \pi
\frac{\omega_{p'}}{q k'},
\end{align}
where the angle is
\begin{align}
\eta_{qk'0}= \frac{1}{2 q k'} \left( -2 \omega_q \omega'+\Sigma_{q k'}
\right),
\end{align}
and
\begin{align}
\label{eq:218}
\Sigma_{q k'}=M_2^2-(\omega'^2-k'^2)-m_\phi^2.
\end{align}
The imaginary part is given by
\begin{align}
\label{imnn''}
\rmIm \left( T \sum_{k_0', h'} \int \frac{\rmd^3 k'}{(2 \pi)^3}
  \D_{N'} \D_{\phi'} \D_{h'} H_- \right)_{N_\phi^{N'}} = & \frac{1}{4 \pi^3}\rmIm \left(
  T \sum_{k_0',h'} \int_0^\infty \rmd k' k'^2 \rmd \eta_{qk'}
  \int_0^\pi \rmd\phi_{qk'} 
\D_{N'} \D_{\phi'} \D_{h'} H_-
\right) \nonumber \\ = & - \frac{1}{16 \pi^2} \sum_{h'} \int \rmd k'
\rmd \phi' \frac{k'}{q \omega_{q'}} Z_{h'} Z_\phi^{N'} (A_\ell^{\phi'}
-A_\ell^0) H_+.
\end{align}

\subsection{Analytic
  Expressions for the $\boldsymbol{\CP}$-Asymmetries}
\label{sec:integr-expr-cp}

\subsubsection*{Vertex cut through
  $\boldsymbol{\{N_2,\phi'\}}$}
\label{sec:vertex-cut:-external}

We get for $N_\ell^{N'}$
\begin{align}
\rmIm(I_V)_{N_\ell^{N'}}=  \frac{M_1 M_2}{16 \pi^2}
\frac{Z_h\omega}{k} \sum_{h h'} \int_0^\infty \rmd q' \int_0^{\pi}
\rmd \phi_{q'} \frac{q'}{\omega_{q'}} Z_{h'}  
 \left[ \left( B_\phi^{\ell'} - B_N^{\ell'} \right) H_-
  + \left( B_\phi^0 - B_N^{\phi'} \right) H_+ \right] \, .
\end{align}
The difference in decay rates reads
\begin{align}
\label{eq:139}
   \gamma(N \to \ell_h \phi)& - \gamma(N \to \barell_h \barphi) = 
 - g_{SU(2)} \rmIm
  \left\{ \left[ \left( \lambda^\dagger \lambda \right)_{12} \right]^2
  \right\} \frac{M_1 M_2}{4 (2 \pi)^5} \nonumber \\
 & \times  \sum_{h'} \int \rmd E \rmd k \rmd q'
  \int_0^\pi \rmd \phi_{q'} k F_{N_ h}^\rmeq Z_h \frac{q'}{k \o_{q'}}
  Z_{h'} \left[ \left( B_\phi^{\ell'} - B_N^{\ell'} \right) H_-
  + \left( B_\phi^0 - B_N^{\phi'} \right) H_+ \right] 
\end{align}
and the $\CP$-asymmetry reads
\begin{align}
\label{eq:136}
\epsilon_h(T) = & - g_{SU(2)} \frac{\rmIm \{[( \lambda^\dagger
  \lambda)_{12}]^2\}}{\gamma(N \to L_h N)} \frac{M_1 M_2}{4 (2 \pi)^5}
\sum_{h'} \int \rmd E \rmd k \rmd q' \int_0^\pi \rmd \phi_{q'} k F_{N_
  h}^\rmeq Z_h \frac{q'}{k \o_{q'}}
Z_{h'} \nonumber \\
& \times \left[ \left( B_\phi^{\ell'} - B_N^{\ell'} \right) H_-
  + \left( B_\phi^0 - B_N^{\phi'} \right) H_+ \right] \nonumber \\
=&- \frac{\rmIm\{[(\lambda^\dagger \lambda)_{12}]^2\}}{g_c
  (\lambda^\dagger \lambda)_{11}} \frac{M_1
  M_2}{4 \pi^2}  \nonumber \\
& \times \frac{\sum_{h'} \int \rmd E \rmd k \rmd k' \int_0^\pi \rmd
  \phi' k F_{N_ h}^\rmeq Z_h \frac{k'}{p \o_{p'}} Z_{h'} [(
  B_\phi^{\ell'} - B_N^{\ell'}) H_- + ( B_\phi^0 - B_N^{\phi'}) H_+
  ]}{\int \rmd E \rmd k k f_N Z_D Z_h(p_0-h p \eta)} \, ,
\end{align}

\subsubsection*{Vertex cut through
  $\boldsymbol{\{N_2,\ell'\}}$}
\label{sec:vertex-cut:-external-4}

For $N_\phi^{N'}$, we get
\begin{align}
\rmIm(I_V)_{N_\phi^{N'}}= \frac{M_1 M_2}{16 \pi^2}
\frac{Z_h\omega}{q} \sum_{h'} \int_0^\infty \rmd k' \int_0^{\pi}
\rmd \phi_{qk'} \frac{k'}{\omega_{q'}} Z_{h'} Z_\phi^{N'}
(A_\ell^{\phi'}-A_\ell^0) H_+ \, .
\end{align}
The difference in decay rates reads
\begin{align}
\label{eq:138}
   \gamma(N \to \ell_h \phi) - \gamma(N \to \barell_h \barphi) = 
& - g_{SU(2)} \rmIm
  \left\{ \left[ \left( \lambda^\dagger \lambda \right)_{12} \right]^2
  \right\} \frac{M_1 M_2}{4 (2 \pi)^5} \nonumber \\
 & \times \sum_{h'} \int \rmd E \rmd k \rmd k'
  \int_0^\pi \rmd \phi_{qk'} k F_{N_ h}^\rmeq Z_h \frac{k'}{q \o_{q'}}Z_\phi^{N'}
  Z_{h'}(A_\ell^{\phi'} -A_\ell^0) H_+ \, .
\end{align}
 The $\CP$-asymmetry reads
\begin{align}
  \epsilon_h(T) = & - g_{SU(2)} \frac{\rmIm \{[( \lambda^\dagger
    \lambda)_{12}]^2\}}{\gamma(N \to L_h N)} \frac{M_1 M_2}{4 (2
    \pi)^5} \sum_{h'} \int \rmd E \rmd k \rmd k' \int_0^\pi \rmd
  \phi_{qk'} k F_{N_ h}^\rmeq Z_h \frac{k'}{q \o_{q'}}Z_\phi^{N'}
  Z_{h'}(A_\ell^{\phi'} -A_\ell^0) H_- \nonumber \\
  =&- \frac{\rmIm\{[(\lambda^\dagger \lambda)_{12}]^2\}}{g_c
    (\lambda^\dagger \lambda)_{11}} \frac{M_1 M_2}{4 \pi^2}
  \frac{\sum_{h'} \int \rmd E \rmd k \rmd k' \int_0^\pi \rmd
    \phi_{qk'} k F_{N_ h}^\rmeq Z_h \frac{k'}{q \o_{q'}}Z_\phi^{N'}
    Z_{h'}(A_\ell^{\phi'} -A_\ell^0) H_-}{\int \rmd E \rmd k k f_N Z_D
    Z_h (p_0-h p \eta)} \, ,
\end{align}
where we integrate over $\phi_{qk'}$ and we take the coordinate system
differently than for $N_N^N$.

\section{Approximation for the One-Mode Approach at High Temperature}
\label{sec:appr-one-mode}

Let us examine the high temperature behaviour of the one-mode approach
by calculating the $\CP$-asymmetry in the matrix elements of a Higgs
boson at rest, where we assume that $M_1, m_\ell \ll m_\phi \ll
M_2$. For simplicity, we calculate the self-energy contribution. The
integral that corresponds to equation~\eqref{eq:137} reads
\begin{align}
  \label{eq:172}
    I_0 I_1^* = 2 T \sum_{k_0'} \int \frac{{\rm d}^3 k'}{(2 \pi)^3} M_2 M_1 \left[
  \Delta_{N'} \Delta_{\phi'} \Delta'_{\ell'} \right]^* K \cdot K' \, .
\end{align}
The part that contributes to the imaginary part of the diagram is
\begin{align}
  \label{eq:173}
  \left. I_0 I_1^* \right|_N^{\phi'} &= 2 \int \frac{{\rm d}^3 k'}{(2
    \pi)^3} M_2 M_1 \Delta_{N'} \frac{1}{4 \omega_{q'} \omega_{k'}}
  B_N^{\phi'} (K \cdot
  K') \, , \nonumber \\
  \rmIm \, B_N^{\phi'}& = - \pi Z_N^{\phi'} \delta(N_N^{\phi'}) = -\pi
  \frac{\omega_{q'}}{k k'} Z_N^{\phi'} \delta(\xi - \xi_0) \, ,
\end{align}
where $\xi \equiv ({\bf k k'})/(k k')$,
\begin{align}
  \label{eq:175}
  \xi_0 = \frac{m_\phi-k'}{k'} \,
\end{align}
and we have neglected $M_1$ and $m_\ell$.
We get
\begin{align}
  \label{eq:176}
  \rmIm(I_0 I_1^*)_N^{\phi'} = - \frac{1}{8 \pi} \frac{\sqrt{x}}{1-x}
  \int_{k}^\infty \rmd k' Z_N^{\phi '} (2 k' - m_\phi) \, ,
\end{align}
where $x \equiv M_2^2/M_1^2$ and $k=m_\phi/2$. For simplicity, we
make the approximation
\begin{align}
  \label{eq:177}
  Z_N^{\phi'}=f_{\phi'}+f_{\ell'} \approx \rme^{- \omega_{q'} \beta} +
  \rme^{-\omega_{k'} \beta} = (1 + \rme^{- k \beta}) \rme^{-k' \beta}
  \, ,
\end{align}
so 
\begin{align}
  \label{eq:178}
    \rmIm(I_0 I_1^*)_N^{\phi'} = - \frac{1}{8 \pi}
    \frac{\sqrt{x}}{1-x} (1+\rme^{-k \beta})
  \int_{k_1'}^\infty \rmd k' \rme^{-k' \beta} (2 k' - m_\phi) \, .
\end{align}
The integral gives
\begin{align}
  \label{eq:179}
      \rmIm(I_0 I_1^*)_N^{\phi'} = - \frac{1}{4 \pi}
    \frac{\sqrt{x}}{1-x} (1+\rme^{-k \beta}) T^2 \rme^{-k \beta} \, .
\end{align}
We parameterise $m_\phi$ as $m_\phi = g_\phi T$ and obtain
\begin{align}
  \label{eq:180}
\Delta \left| \mathcal{M} \right|^2 \equiv
| \mathcal{M} |^2- | \widetilde{\mathcal{M}} |^2
= - 8 \rmIm \lambda_{CP} \rmIm(I_0 I_1^*)= 
2 \frac{\rmIm \lambda_{CP}}{\pi}
    \frac{\sqrt{x}}{1-x} T^2 \rme^{-g_\phi/2} (1 +\rme^{-g_\phi/2}) \, .
\end{align}
Using the expression
\begin{align}
  \label{eq:182}
  | \mathcal{M}_{\rm tot} |^2 = 4 (\lambda^\dagger \lambda)_{11} K \cdot
  P = 2 (\lambda^\dagger \lambda)_{11} g_\phi^2 T^2
\end{align}
we arrive at
\begin{align}
  \label{eq:181}
  \epsilon = \frac{\Delta | \mathcal{M} |^2}{|\mathcal{M}_{\rm
      tot}|^2} =  \frac{\rmIm \lambda_{CP}}{(\lambda^\dagger
    \lambda)_{11}} \frac{1}{\pi} \frac{\sqrt{x}}{1-x} \frac{1}{g_\phi^2}
  \rme^{-g_\phi/2} (1+ \rme^{-g_\phi/2}) = \frac{8}{g_\phi^2}
  \rme^{-g_\phi/2} (1 + \rme^{-g_\phi/2})
  \epsilon_0 .
\end{align}
Assuming that $g_\phi \ll 1$, we get
\begin{align}
  \label{eq:184}
  \frac{\epsilon_{\rm rf}^{T \gg M_1}}{\epsilon_0} \approx \frac{32}{g_\phi^2}
\end{align}
Taking $g_\phi = m_\phi/T \approx 0.42$ for $T=10^{12} \, {\rm GeV}$
and using the more accurate term in equation \eqref{eq:181}, we get
$\epsilon/\epsilon_0 \approx 70$, while we get $\epsilon/\epsilon_0
\approx 90$ for equation~\eqref{eq:184}. We view this result as a
rough approximation of the value of the $\CP$-asymmetry in Higgs boson
decays at high temperature. Both our approximation and the numerical
solution of the exact expression in section~\ref{sec:results} give a
factor of 100 difference to the $\CP$-asymmetry in vacuum.

\bibliographystyle{JHEPMS}
\bibliography{../dissertation/main}

\end{document}